\documentclass[10pt]{article}
\pdfoutput=1

\usepackage[utf8]{inputenc}
\usepackage[left=2.55cm, right=2.55cm, top=2.55cm, bottom=2.55cm]{geometry}
\usepackage{amsmath,amssymb,amsbsy}
\usepackage{slashed}
\usepackage{xcolor}
\usepackage{graphicx}
\usepackage{url}
\usepackage{cancel}
\usepackage{cite}
\usepackage[colorlinks=true,allcolors=blue,pdfborder={0 0 0},linktocpage=false,pdfencoding=auto]{hyperref}
\usepackage{tabularx,booktabs}
\usepackage{multicol}
\usepackage{feynmp}
\usepackage{units}
\usepackage{xspace}
\usepackage[labelfont=bf]{caption}
\usepackage[section]{placeins}
\usepackage{subcaption}
\usepackage{soul}
\usepackage{bbold}
\usepackage{parskip}
\usepackage{float}
\usepackage{tabulary}
\usepackage{ulem}

\definecolor{darkred}{rgb}{0.6,0,0}
\definecolor{darkpurple}{rgb}{0.5,0,0.5}
\definecolor{darkgreen}{rgb}{0.0,0.7,0}
\newcommand{\code}[1]{\texttt{#1}}
\newcommand{\beqn}{\begin{eqnarray}}
\newcommand{\eeqn}{\end{eqnarray}}

\begin{document}
\author{Amin Aboubrahim$^a$\footnote{\href{mailto:aabouibr@uni-muenster.de}{aabouibr@uni-muenster.de}}~, Lutz Althueser$^b$\footnote{\href{mailto:l.althueser@uni-muenster.de}{l.althueser@uni-muenster.de}}~, Michael Klasen$^a$\footnote{\href{mailto:michael.klasen@uni-muenster.de}{michael.klasen@uni-muenster.de}}~, Pran Nath$^c$\footnote{\href{mailto:p.nath@northeastern.edu}{p.nath@northeastern.edu}}~~ \\ and Christian Weinheimer$^b$\footnote{\href{mailto:weinheim@uni-muenster.de}{weinheim@uni-muenster.de}}  \\~\\
$^{a}$\textit{\normalsize Institut f\"ur Theoretische Physik, Westf\"alische Wilhelms-Universit\"at M\"unster,} \\
\textit{\normalsize Wilhelm-Klemm-Stra{\ss}e 9, 48149 M\"unster, Germany} \\
$^{b}$\textit{\normalsize Institut f\"ur Kernphysik, Westf\"alische Wilhelms-Universit\"at M\"unster,} \\
\textit{\normalsize Wilhelm-Klemm-Stra{\ss}e 9, 48149 M\"unster, Germany} \\
$^{c}$\textit{\normalsize Department of Physics, Northeastern University,} \\
\textit{\normalsize 111 Forsyth Street, Boston, MA 02115-5000, U.S.A}
}

\title{\vspace{-2cm}\begin{flushright}
{\small MS-TP-22-21}
\end{flushright}
\vspace{1cm}
\Large \bf
Annual modulation of event rate and electron recoil energy in inelastic scattering direct detection experiments
 \vspace{0.5cm}}

\date{}
\maketitle

\begin{abstract}

In 2020 the XENON1T experiment observed an excess of events with an electron recoil energy $E_R$ in the range of $2\,$--$\,3\,$keV. Such an excess can arise from a variety of sources such as solar axions or a neutrino magnetic moment, but also from inelastic scattering of dark matter off the xenon atoms. The recoil energy of the electron then depends on the mass difference of the dark particles. In this paper we show that the annual modulation of both the event rate and the electron recoil energy provide important additional information that allows to distinguish among different theoretical explanations of the signal. To this end, we first extend the formalism of annual modulation to electronic recoils, inelastic dark matter scattering and the electron recoil energy. We then study a concrete theoretical model with two Dirac fermions and a dark photon. We take into account all relevant cosmological and experimental constraints on this model and apply it to the XENON1T and and XENONnT experiments with realistic detection thresholds, efficiencies and energy resolutions, fitting the main physical parameters of the model, i.e. the mass splitting and the electron scattering cross section. The discriminatory power of the additional information from the annual modulation of both the signal rate and the electron recoil energy is then demonstrated for XENONnT with a simplified model based on these main physical parameters. This more sensitive procedure compared to time-only modulation analyses can also serve as a template for other theoretical models with different dark matter candidates, mediators and cosmology. For the $U(1)$ model with two Dirac fermions fitting the XENON1T excess and the experimental conditions of XENONnT, taking into account the annual variation of the signal rate and recoil energy allows for a faster and more precise determination of the free model parameters.

\end{abstract}

\numberwithin{equation}{section}

\newpage

{  \hrule height 0.4mm \hypersetup{colorlinks=black,linktocpage=true} \tableofcontents
\vspace{0.5cm}
 \hrule height 0.4mm}

\section{Introduction\label{sec1}}

Recently the XENON Collaboration has analyzed 0.65 tonne-years of XENON1T detector data in the electron recoil energy range of $1\,$--$\,30\,$keV \cite{Aprile:2020tmw}. An excess of electron recoil events over the background was observed in the energy range of $2\,$--$\,3\,$keV. The collaboration analyzed several possible sources to account for the excess which included axion couplings and a neutrino magnetic moment as possible sources. These, however, appear to be in tension with stellar constraints. The possibility that the excess could be due to contamination by traces of tritium in xenon at the level of $(6.2\pm 2.0)\times 10^{-25}$ mol/mol was also investigated, but this possibility was neither confirmed nor excluded by the collaboration. There has been a large number of theory papers following the XENON1T results (see ref.~\cite{Aboubrahim:2020iwb} and the references therein). In this work we wish to focus on the possibility of observing annual modulations in the XENON1T and XENONnT detectors. In addition to analyzing the annual modulation in the event rates, we propose that modulations in electron recoil energy would be an additional important signal for dark matter (DM). We show that the amplitude of the recoil energy fluctuation can provide a direct measurement of the DM mass. 
 
Annual modulations arise due to the relative motion of the Earth around the Sun. One of the key advantages of this method to detect DM is that the background from known sources is largely insensitive to modulation and this reduction in the background makes annual modulation of DM event rates an attractive possibility (for a comprehensive review of annual modulation see ref.~\cite{Freese:2012xd}). The utility of the annual modulation for detecting DM has a well established history. 
DAMA and DAMA-LIBRA have reported a strong modulation signal for many years~\cite{Bernabei:2003za,Bernabei:2008yh,Bernabei:2010mq,Bernabei:2018jrt}, but these results are disputed, partly because they have not been confirmed by other experiments. CoGENT~\cite{Aalseth:2011wp} and CRESST-II \cite{Angloher:2011uu} have also claimed modulation signals, although the results are uncorroborated, and COSINE-100~\cite{COSINE-100:2021zqh} performed a modulation analysis with an inconclusive result. Annual modulation has also been considered in various other DM experimental searches~\cite{Adhikari:2019off,Coarasa:2018qzs,Kobayashi:2018jky,deSouza:2016fxg,Aalseth:2014eft,Bernabei:2014jba,Cherwinka:2011ij,Aalseth:2008rx} and theoretical considerations \cite{Arnowitt:1999gq,DelNobile:2015rmp,Witte:2016ydc,TuckerSmith:2001hy,Barello:2014uda,Bozorgnia:2013hsa,Bramante:2016rdh}.

In this work we discuss the annual modulation of the DM signal arising from the inelastic scattering process $e+X_2 \to e+X_1$ where the dark particle $X_2$ down scatters from an electron to a dark particle $X_1$ which is less massive than $X_2$ giving an excess energy to the recoiling electron. Our analysis is in the framework of ref.~\cite{Aboubrahim:2020iwb} which can explain the excess seen in ref.~\cite{Aprile:2020tmw}. We extend the analysis of annual modulation of the event rate to electronic recoils and inelastic scattering and point out that also the electron recoil energy is modulated. A measurement of this modulation can then lead to a direct determination of the DM mass.

The outline of the paper is as follows: In section \ref{sec2} we present a derivation of the modulation in event rate and event rate for electronic recoils and inelastic scattering. In section \ref{sec3} we introduce the specific model used in this analysis and calculate the corresponding DM relic density. The XENON1T and XENONnT experiments are discussed in section \ref{sec4}, together with the fits of our model to the excess in electron recoil energy observed in XENON1T. In section \ref{sec5} we then propose a simplified annual modulation model, which can also serve as a template for other DM models. Our analysis methods, sensitivity studies and numerical results are given in section \ref{sec6}.  Conclusions are given in section \ref{sec7}. Further details regarding the model, DM inelastic scattering and calculations of the event rate are given in appendices~\ref{app:A},~\ref{app:B} and~\ref{app:C}. 

\section{Modulation in recoil energy and event rate\label{sec2}}

We discuss first a model-independent inelastic scattering process. Let us consider two dark particles $X_1$ and $X_2$ with masses $m_1$ and $m_2$ and a small mass splitting $\Delta m=m_2-m_1>0$. In this case, when the dark particle $X_2$ hits a bound electron in a xenon atom, it  produces, in an exothermic process, a recoil electron with excess energy, i.e.,  
\begin{align}
e+ X_2\to e+ X_1, 
\label{2.1} 
\end{align}
where the final electron receives an extra boost in energy from the mass difference $\Delta m$. Now the recoil energy of the scattered electron can exhibit an annual modulation effect which arises due to the motion of the Earth around the Sun. The velocity of the Earth with respect to the DM halo is given by
\begin{equation}
v_{EH}(t)=\sqrt{v^2_{ES}+v^2_{SH}+2v_{ES}v_{SH}\cos\gamma\cos(2\pi t-\phi)},
\label{relv}
\end{equation}
where $\phi$ is the phase on June 2nd so that $\phi=2.61$ (with $\phi=0$ being on January 1) and $\cos\gamma\simeq 0.5$. The speed of the Earth relative to the Sun is $v_{ES}=29.8$ km/s and the speed of the Sun relative to the halo is $v_{SH}=233$ km/s.
The electron recoil energy also has a velocity-dependent effect which is subject to 
annual modulation, because of Eq.~(\ref{relv}), i.e., 
\begin{align}
 E_R(t)&= \Delta m  + \frac{1}{2} m_2 v_{EH}^2(t), 
 \label{ER}
\end{align}
where $v_{EH}(t)$ is given by Eq.~(\ref{relv}).
The variation in $E_R$ over the course of one year is  then given by
\begin{align}
 \Delta E_R&=   2 m_2 v_{SH} v_{ES} \cos\gamma\,.
 \label{deltaER}
 \end{align}
Remarkably, Eq.~(\ref{deltaER}) is largely a model-independent result with the only model
dependence coming from the mass $m_2$ of the dark particle $X_2$. Of course the
 resulting event rates will be model-dependent and it is of interest to compute the size of the recoil energy modulation $\Delta E_R$ in concrete particle physics models which accommodate the 
  excess seen by the XENON1T experiment and to asses if such a modulation can be seen in future XENONnT experiment. To give a concrete example, using Eq.~(\ref{ER}), we find a variation of 1$-$3\% in the recoil energy over the course of one year for a DM mass in the range 0.3 GeV$-$1.0 GeV. However, larger DM masses compatible with the relic density constraint could achieve larger variations.  Thus, a 10 GeV DM particle would lead to a 30\% variation in the recoil energy.

Next, we calculate the basic equations needed in our analysis to study the annual modulations in the event rate as well as in the recoil energy. It is a generalization of the analysis in ref.~\cite{Aboubrahim:2020iwb} where the time component is added to model the effect of annual modulations.

In the inelastic scattering of Eq.~(\ref{2.1}), the incident DM has mass $m_2$ and momentum $\mathbf{p}$ and the scattered DM has mass $m_1$ and momentum $\mathbf{p}-\mathbf{q}$, where $\mathbf{q}$ is the momentum transfer. Energy conservation gives
\begin{align}
E_R-\Delta m\left[1-\frac{1}{2}\left(\frac{m_2}{m_1}\right)v^2\right]=\left(\frac{m_2}{m_1}\right)\mathbf{v}\cdot \mathbf{q} -\frac{q^2}{2\mu_{1N}},
\end{align}
where the DM-nucleus reduced mass is $\mu_{1N}=m_1 m_N/(m_1+m_N)$. Taking $m_1\approx m_2\approx m_D$, the range of momentum transfer is given by
\begin{equation}
    q_{\pm}= 
\begin{cases}
    m_D v\pm\sqrt{m_D^2 v^2-2m_D(E_R-\Delta m)},& \text{for } E_R>\Delta m,\\
    \pm m_D v+\sqrt{m_D^2 v^2-2m_D(E_R-\Delta m)}, & \text{for } E_R<\Delta m.
\end{cases}
\end{equation}
The velocity-averaged differential cross-section for inelastic DM scattering is
\begin{equation}
\frac{d\langle\sigma v\rangle}{d E'}=\frac{\bar\sigma_e}{2m_e}\eta(v_{\rm min},t)\int_{q_{-}}^{q_{+}} dq\,a_0^2 q K(E',q),
\label{K-int}
\end{equation}
where the Bohr radius $a_0=1/(\alpha_{\rm em} m_e)$ $(\alpha_{\rm em}\simeq 1/137)$ and $K(E,q)$ is the atomic factorization factor. The quantity $\eta(v_{\rm min},t)$ holds all astrophysical information and is given by
\begin{equation}
\eta(v_{\rm min},t)=\int_{v_{\rm min}}^{v_{\rm max}}\frac{f(\textbf{v}+\textbf{v}_{EH})}{v}d^3 v,
\label{etaf}
\end{equation} 
where the Boltzmann velocity distribution of DM, $f(\textbf{v}+\textbf{v}_{EH})$, is written after a Galilean boost $\textbf{v}\longrightarrow \textbf{v}+\textbf{v}_{EH}$ from the DM rest frame to the lab frame, with $\textbf{v}_{EH}$ being the velocity of the Earth with respect to the DM halo, so that $\textbf{v}_{EH}(t)=\textbf{v}_{ES}(t)+\textbf{v}_{SH}(t)$. Here $\textbf{v}_{ES}$ is the velocity of the Earth with respect to the Sun and $\textbf{v}_{SH}$ is the velocity of the Sun with respect to the halo. 
Therefore we have
\begin{equation}
f(\textbf{v}+\textbf{v}_{EH})=\frac{1}{N_{\rm esc}}(\pi v_0^2)^{-3/2} e^{-|\textbf{v}+\textbf{v}_{EH}|^2/v_0^2},
\label{maxwell}
\end{equation}
where the normalization factor is given by
\begin{equation}
N_{\rm esc}=\text{Erf}\left(\frac{v_{\rm esc}}{v_0}\right)-\frac{2v_{\rm esc}}{\sqrt{\pi}v_0}e^{-v^2_{\rm esc}/v_0^2},
\end{equation}
with $v_0=220$ km/s being the most probable velocity and $v_{\rm esc}=544$ km/s the escape velocity. The minimum velocity is taken to be
\begin{equation}
v_{\rm min}(E')=\sqrt{\text{max}\left\{\frac{2}{m_2}(E'-\Delta m),0\right\}},   
\end{equation}
and $v_{\rm max}=v_{EH}(t)+v_{\rm esc}$.
Using Eqs.~(\ref{maxwell}) and~(\ref{relv}), we evaluate Eq.~(\ref{etaf}) to get
\begin{align}
\eta(E',t)&=\frac{1}{2 N_{\rm esc} v_{EH}(t)}\left[\text{Erf}\left(\frac{v_{EH}(t)+v_{\rm min}(E')}{v_0}\right)+\text{Erf}\left(\frac{v_{EH}(t)-v_{\rm min}(E')}{v_0}\right)-\frac{4v_{EH}(t)}{\sqrt{\pi}v_0}e^{-v^2_{\rm esc}/v^2_0}\right] \nonumber \\
&\approx \frac{1}{2 v_{\textrm{EH}}(t)}\left[ \text{Erf}\left(\frac{v_{\textrm{EH}}(t) + v_\textrm{min}(E')}{v_0}\right) + \text{Erf}\left(\frac{v_{\textrm{EH}}(t) - v_{\textrm{min}}(E')}{v_0}\right) \right],
\end{align}
where the approximation in the last step is in the limit $v_{\rm esc}\to\infty$ which implies $N_{\rm esc}\to 1$.

In Eq.~(\ref{K-int}), the integral on $dq$, i.e., 
\begin{equation}
K'(E',t)\equiv \int_{q_{-}}^{q_{+}} dq\,a_0^2 q K(E',q),
\end{equation}
can be parameterized by
\begin{equation}
K'(E',t)=\frac{a_0^2}{w^2\sqrt{\pi}}e^{-(E'-E_R(t))^2/w^2},
\end{equation}
where $E_R(t)$ is given by Eq.~(\ref{ER}) and a fitting to the integrated atomic ionization factor for xenon~\cite{Roberts:2019chv} gives $w\simeq 0.046$. Due to the modulating nature of $E_R(t)$, this function exhibits a narrow peak whose position is slightly shifted away from $\Delta m$.

The resulting energy spectrum and event rates have to be corrected for detector effects such as the detection threshold, detection efficiency and energy resolution. Here we follow a similar approach as the XENON experiment~\cite{Aprile:2020tmw}. The detector resolution is given by
\begin{equation}
\sigma_E(E')=a\sqrt{E'}+b ~E', \label{energy_resolution}
\end{equation}
with the detector-dependent parameters $a$ and $b$ as given in section~\ref{sec4}. The DM detection rate is given by
\begin{align}
\frac{dR}{dE\,dt}(E,t)&=n_{\rm Xe}\frac{\rho_2}{m_2}\int\frac{d\langle\sigma v\rangle}{d E_v}R_S(E,E_v)dE_v \nonumber \\
&=n_{\textrm{Xe}} \frac{\rho_2}{m_2} \frac{\bar\sigma_e}{2m_e} \int  \eta(E_v,t) \, K'(E_v, t) \, R_S(E, E_v)\,dE_v\,,
\label{dRdE}
\end{align}
where $n_{\rm Xe}\simeq 4.2\times 10^{27}/$ton is the number of xenon atoms in the detector and $\rho_2\simeq 0.15$ GeV/cm$^3$ assuming that $X_2$ makes half the amount of the observed relic density (see below). The integral in Eq.~(\ref{dRdE}) is carried over the entire energy spectrum $E_v$. We assume the resolution function is a Gaussian of the form
\begin{equation}
R_S(E, E_v) = \frac{1}{\sqrt{2\pi}\sigma_E(E_v)}\exp\left[-\frac{(E-E_v)^2}{2 \sigma^2_E(E_v)}\right]~\alpha(E),
\end{equation}
where $\alpha(E)$ is the detector efficiency discussed in section \ref{sec4}.

\section{Model-dependent analysis}\label{sec3}  
  
\subsection{The model}

In this section we present the model we will use in the analysis of the annual modulations of the event rate as well as the recoil energy. The proposed model consists of two
Dirac particles in the hidden sector with an extra $U(1)$ gauge symmetry which interact with
the visible sector via gauge interactions involving a dark photon $\gamma'$ and the $Z$
boson. The extra $U(1)$ has kinetic mixing~\cite{Holdom:1985ag}  
with the visible sector and gains mass via the 
Stueckelberg mechanism~\cite{Kors:2004dx,Cheung:2007ut,Feldman:2007wj,Aboubrahim:2020lnr}. 
In the basis where the kinetic and mass terms are canonically normalized, the interactions of the dark photon $\gamma'$ with mass $m_{\gamma'}$ and of the $Z$ boson with the dark Dirac
fermions $D_1'$ with mass $m_1$ and $D_2'$ with mass $m_2 > m_1$ are given by 
\begin{equation}
\begin{aligned}
-\mathcal{L}^{\rm int}_D&= \frac{1}{2} (Q_1+ Q_2) \left( \bar D_1'\gamma^\mu D_1' 
+ \bar D_2'\gamma^\mu D_2'\right) (g^{\gamma'}_X A_\mu^{\gamma'}+ g^Z_X Z_\mu)\\
&+\frac{1}{2} (Q_1-Q_2) (\bar D_1' \gamma^{\mu}D_2' + \bar D_2' \gamma^{\mu}D_1')
(g^{\gamma'}_X A_\mu^{\gamma'}+ g^Z_X Z_\mu),
\end{aligned}
\label{2-d1d2-2} 
\end{equation}
where $g^{\gamma'}_X$ and $g^Z_X$ are the gauge coupling constants and $Q_1,Q_2$ are the $U(1)_X$ charges defined in Appendix~\ref{app:A}, where the details of the deduction of Eq.~(\ref{2-d1d2-2}) 
are given.  Here 
  from Eq.~(\ref{2-d1d2-2}) and Eq.~(\ref{SMLag})   
   we find that the dark photon has couplings
  with both the visible sector and the hidden sector. 
  Thus the dark photon couples with quarks and leptons in the visible sector 
  and  with  $D_1', D_2'$ in the dark sector. 
We note here that the  DM-nucleon elastic scattering cross-section against a nucleus with mass number $A$ and proton number $Z$ is given by 
\begin{equation}
\sigma_{\text{DN}}=\frac{g_X^2 (Q_1+Q_2)^2 g_2^2 v_f^{\prime 2}}{16\pi\cos^2\theta}\frac{\mu^2_{DN}}{m^4_{\gamma'}}\left(\frac{Z}{A}\right)^2,
\end{equation} 
where $\theta$ is the weak mixing angle in the extended model (see Appendix~\ref{app:A} for further details), $\mu_{DN}$ is the DM-nucleon reduced mass, and $v_f'$ enters in the coupling of 
the dark photon with the visible sector as given in Appendix~\ref{app:A}.
 In the analysis we choose $Q_1+Q_2=0$ and thus the DM-nucleon cross section vanishes
for this case. However, the DM can also scatter off a bound electron in the xenon atom
by elastic and inelastic scattering. The elastic scattering of DM with a bound electron in the xenon atom can deliver only a few eV to the electron which is not sufficient to explain the XENON1T excess.
However, an inelastic exothermic down-scattering can impart a recoil energy to the electron equivalent to the mass difference between the incoming and outgoing DM particles. 
The model considered here  allows for the desired small mass splitting between the two Dirac fermions $D'_1$ and $D'_2$ so that the heavier fermion $D'_2$ down-scatters to $D'_1$. 
Under the assumption that the dark photon mass is much greater than the momentum transfer,
one has the following inelastic scattering cross-section of the process $D_2'(\vec{p}_2)+e(\vec{p}_1)\to D_1'(\vec{p}_4)+e'(\vec{p}_3)$,
\begin{equation}
\bar{\sigma}_e\simeq\frac{\bar{g}_X^2 g_2^2}{4\pi\cos^2\theta}\frac{\mu^2_{De}}{m^4_{\gamma'}}v_f^{\prime 2}.
\label{inelastic-cross}
\end{equation}
Details of the analysis are given in Appendix~\ref{app:B}.
 One finds that the cross-section depends on the gauge coupling in the dark sector and on the kinetic mixing which enters in the expression through $v'_f$. Such quantities are constrained by experiment.

\subsection{Dark matter relic density}

The size of the coupling between DM and SM particles in this model is determined by the kinetic mixing parameter $\delta$ and the dark coupling $g_X$. For a fixed $g_X\sim\mathcal{O}(10^{-2})$, a kinetic mixing $\delta\lesssim 10^{-7}$ implies that the visible and hidden sectors are feebly coupled. Therefore there is no a priori reason to assume an initial thermal distribution of DM in the early universe. DM particles may have been negligible in the early universe but despite the feeble couplings with the SM, the DM population can still grow from SM particle annihilation to DM particles through the $Z$ and $\gamma'$ portals. As the number density of dark particle species increases the universe cools to the point where the injection processes become inefficient and DM freezes-in. The freeze-in mechanism~\cite{Hall:2009bx} can be seen as the opposite of the standard freeze-out mechanism where the latter assumes a weak scale coupling between the DM particles and the SM.  Much larger values of the kinetic mixing, for e.g. $\delta\gtrsim 10^{-4}$, can efficiently produce a thermal DM distribution and the resulting relic density will be completely determined by the freeze-out mechanism. Intermediate values of $\delta$ can bring about more complicated dynamics where both the freeze-in and freeze-out mechanisms play an essential role in determining the DM relic density. This range of $\delta$ has been explored in previous works, see refs.~\cite{Aboubrahim:2022gjb,Aboubrahim:2021ohe,Aboubrahim:2021dei,Aboubrahim:2021ycj,Aboubrahim:2020lnr}. Furthermore, the dark sector and the visible sector can be at different initial temperatures. The evolution of the dark species' number densities is coupled to the dark and visible temperatures. A generalization of this scheme to $n$ dark sectors is given in ref.~\cite{Aboubrahim:2022bzk}.

In this work we assume a negligible initial abundance of the dark sector species and that the visible and dark sectors are at the same temperature. We focus on the evolution of the DM number density which is produced by $2\to 2$ processes involving SM annihilation to DM. The dark photon is produced via $2\to 2$ and $2\to 1$ processes from the SM. It will eventually decay back to the SM and so it's final yield has no contribution to the relic density. One should note that $D\bar{D}\to\gamma'\gamma'$ plays a role in DM depletion as well as in bringing the dark sector species into thermal equilibrium. 

To calculate the DM relic density of $D'_1$ and $D'_2$, we assume $m_{1}\simeq m_{2}\simeq m_D$ and write only one Boltzmann equation for the dark fermion. In this limit, we have
\begin{equation}
\frac{dY_D}{dx}\approx -1.32 M_{\rm Pl}\frac{h_{\rm eff}(T)}{g^{1/2}_{\rm eff}(T)}\frac{m_D}{x^2}\left(-\langle\sigma v\rangle_{D\bar{D}\to i\bar{i}} Y_D^{\rm eq^2} +\langle\sigma v\rangle_{D\bar{D}\to\gamma'\gamma'} Y^2_D\right),
\label{yield}
\end{equation} 
where $Y_D=n/s$ is the comoving number density (or yield) of DM, $h_{\rm eff}$ and $g_{\rm eff}$ are the entropy and energy density numbers of degrees of freedom,  $M_{\rm Pl}$ is the reduced Planck mass
($M_{\rm Pl} \sim 2.4 \times 10^{16}$ GeV) and $x=m_D/T$. The first term on the right-hand-side of Eq.~(\ref{yield}) is responsible for DM production from the SM while the second term depletes the DM through annihilation to dark photons. Here the thermally averaged cross-section is given by
\begin{equation}
\langle\sigma v\rangle^{D\bar{D}\to ab}(x)=\frac{x}{8 m^5_D K^2_2(x)}\int_{4m_D^2}^{\infty} ds ~\sigma(s) \sqrt{s}\, (s-4m_D^2)K_1\left(\frac{\sqrt{s}}{m_D}x\right),
\end{equation}
while the equilibrium yield is given by
\begin{equation}
Y_D^{\rm eq}(x)=\frac{45}{4\pi^4}\frac{g_D}{h_{\rm eff}(T)} x^2 K_2(x).
\end{equation}
Here $g_D$ is the dark fermion number of degrees of freedom, $K_1$ and $K_2$ are the modified second order Bessel functions of degree one and two. The Boltzmann equation Eq.~(\ref{yield}) 
is solved numerically to determine the yield at the present time $Y_{\infty}$ which gives us the relic density
\begin{equation}
\Omega h^2=\frac{m_D Y_{\infty}s_0 h^2}{\rho_c},
\end{equation}
where $s_0$ is today's entropy density, $\rho_c$ is the critical density and $h=0.678$ denotes the present Hubble expansion rate in units of 100 km s$^{-1}$ Mpc$^{-1}$. In our analysis we require that the DM relic density satisfies, within theoretical uncertainties, the experimental value from the Planck Collaboration~\cite{Aghanim:2018eyx}
\begin{equation}
(\Omega h^2)_{\rm Planck}=0.1198\pm 0.0012.
\end{equation}
We assume that the DM relic density is shared equally between $D'_1$ and $D'_2$. A priori, the ratio of the number densities of $D'_1$ and $D'_2$ is Boltzmann suppressed, i.e.,  $n_{2}/n_{1}\sim\exp(-\Delta m/T_c)$, where  $T_c$ is the temperature below which conversion processes shut off. However, since the cross section of the conversion process, $D'_2\bar{D}'_2\longleftrightarrow D'_1\bar{D}'_1$, is proportional to $(Q_1+Q_2)^2$, which we take to be zero, such a process does not exist. Apart from conversion, the dark fermion $D_2'$ can only decay to  $D_1'$. The only decay channel would be $D_2'\to D_1'\bar \nu\nu$, which is suppressed by the dark photon coupling to the neutrinos which is proportional to the kinetic mixing, and further the decay is phase-space suppressed because
of the small mass gap $\Delta m\sim\mathcal{O}$(keV).  To lowest order in $\Delta m$, the decay width is given by
\begin{equation}
\Gamma_{D_2'\to D_1'\nu\bar\nu}\simeq\frac{x_{\nu}^2(\Delta m)^5}{40\pi^3 m^4_{\gamma'}},
\end{equation}
where for a small gauge kinetic mixing, $g_X^{\gamma'}\approx g_X$ and
\begin{equation}
x_{\nu}\sim g_X g_Y(Q_1-Q_2)\left(\frac{m_{\gamma'}}{m_Z}\right)\delta.
\end{equation}
In this work we are interested in $\Delta m\sim 3$ keV and $\delta\sim 10^{-5}$ which results in a decay lifetime of $D_2'$ of order $10^{13}$ years. This implies that $D_2'$ is stable over the lifetime of the universe. Thus, in this model dark matter is equally constituted of two dark fermions with essentially
degenerate masses which are of order 1 GeV.

\section{The XENON1T and XENONnT experiments and fits to the electron recoil excess\label{sec4}}

The expected number of detected electron recoil events for the theoretical model described in section \ref{sec3} can be calculated for the XENON1T and XENONnT experiments by adding the actual and predicted detector response and scaling with the total exposure.

\begin{figure}[H]
\centering
\includegraphics[width=0.99\textwidth]{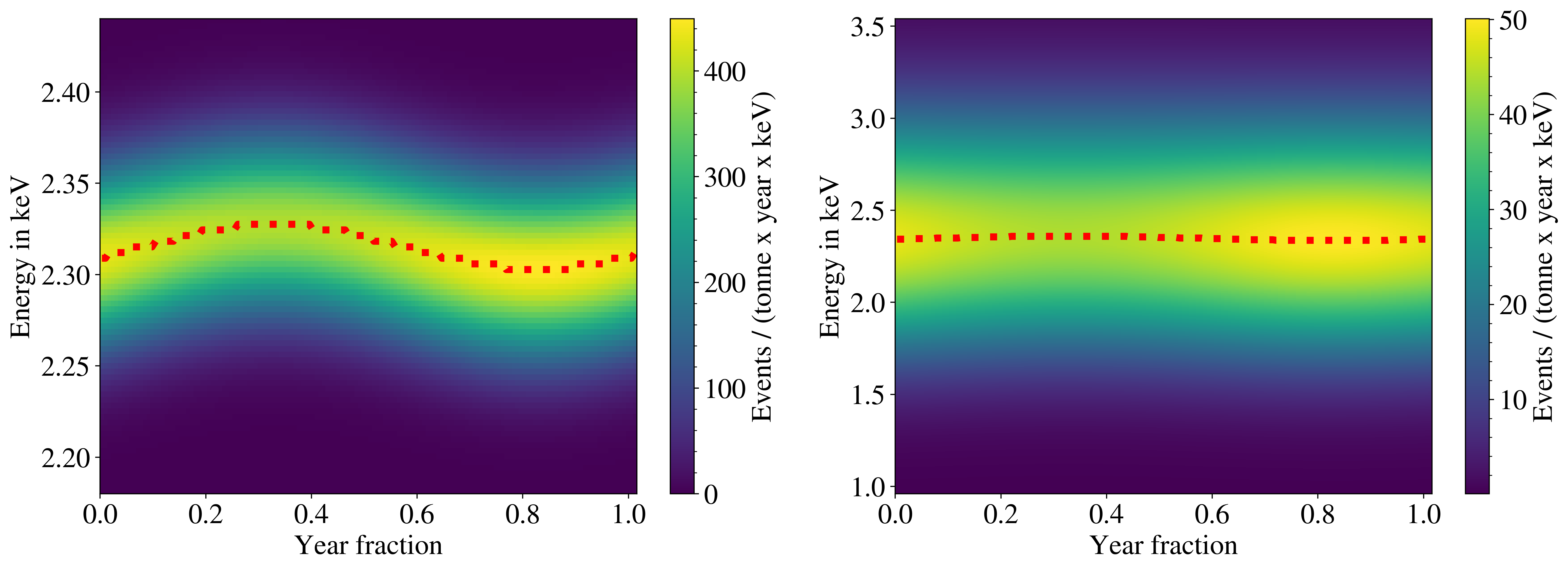}
\caption{Electronic recoil energy spectrum and detected event rates given by $\frac{dR}{dE\,dt}(E,t)$ without (left) and with (right) detector corrections such as efficiencies and energy resolution for the SR1 of the XENON1T experiment as modeled within this framework. The model parameters are set to $m_2 = 0.3\,\textrm{GeV}$, $\Delta m = 2.23\,\textrm{keV}$ and $\bar{\sigma}_e = 3.36\times10^{-44}$ cm$^2$. The dashed red line indicates the maximum event rate over time. Please note that the vertical energy scales of the left and right panels differ significantly.}
\label{model_df_1T_compare}
\end{figure}

The XENON1T experiment performed data taking from February 2017 to February 2018 during science run 1 (SR1)~\cite{Aprile:2020tmw} which has a total duration of 372 days. Data taking breaks and mostly periodic calibrations interrupted the background data taking resulting in an effective live time of 226.9 days. Those interruptions in the background data taking lasted several hours to a few days as reported in ref.~\cite{XENON:2019ykp}. The exact time ranges of good data taking are not known so that we opted to use an effective approach within this framework. The fraction of good data taking over the full SR1 is determined to be 0.61 which is used to scale the detector live time of each day during the modeled science run. This approach neglects the impact of longer interruptions in the background data taking by distributing them uniformly and reduces the impact on the sensitivity to modulations. The result of the actual XENON1T experiment could exhibit a bias in a way that the detector was not taking data during periods in which the recoil rate as determined from the theoretical model was at the maximum. This framework is therefore following a more conservative approach. A publication of the data taking time ranges for the XENON1T experiment would allow for a more robust estimation of the sensitivity. The total xenon mass of the XENON1T experiment that was used after Fiducial Volume (FV) considerations during SR1 is 1042 kg~\cite{Aprile:2020tmw}. Electronic recoil energy spectra are corrected for detector effects such as efficiencies and energy resolution. The selection and detection efficiencies given as $\alpha(E)$ are taken from Fig.~2 of ref.~\cite{Aprile:2020tmw} while the energy resolution is modeled by Eq.~(\ref{energy_resolution}) using $a=(0.310\pm 0.004)\sqrt{\text{keV}}$ and $b=0.0037\pm 0.0003$~\cite{Aprile:2020tmw}. The resulting expectation from the theoretical annual modulation model from section~\ref{sec3} for the XENON1T experiment is shown in Fig.~\ref{model_df_1T_compare}. Here, $m_2 = 0.3\,\textrm{GeV}$, $\Delta m = 2.23\,\textrm{keV}$ and $\bar{\sigma}_e = 3.36\times10^{-44}$ cm$^2$ are used to illustrate an annual modulation signal strength similar to the excess of electronic recoil events observed in ref.~\cite{Aprile:2020tmw}. An annual modulation in the total event rate $\frac{dR}{dE\,dt}(E,t)$ over time as well as in the mean recoil energy can be observed.

\begin{figure}[H]
\centering
\includegraphics[width=0.99\textwidth]{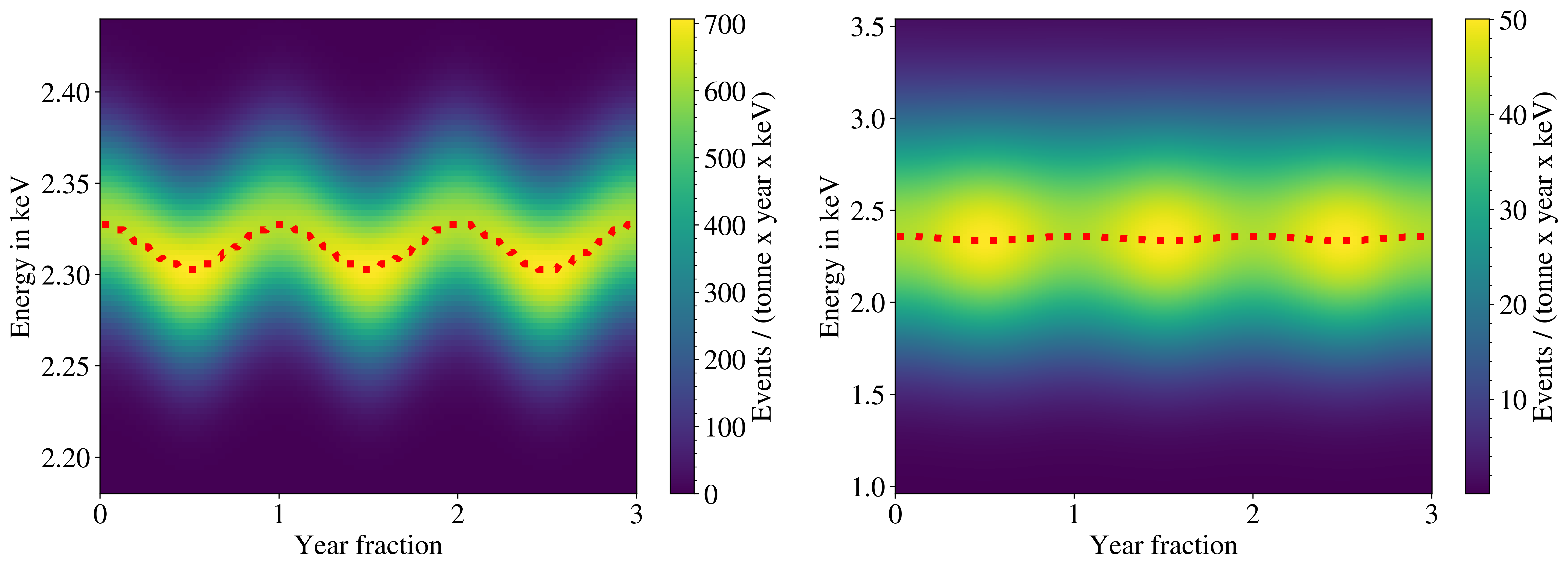}
\caption{Electronic recoil energy spectrum and detected event rates given by $\frac{dR}{dE\,dt}(E,t)$ without (left) and with (right) detector corrections such as efficiencies and energy resolution for the XENONnT experiment after three years of data taking with an effective live time fraction of $0.61$ as modeled within this framework. The model parameters are set to $m_2 = 0.3\,\textrm{GeV}$, $\Delta m = 2.23\,\textrm{keV}$ and $\bar{\sigma}_e = 3.36\times10^{-44}$ cm$^2$. The dashed red line indicates the maximum event rate over time. Please note that the vertical energy scales of the left and the right panels differ significantly.}
\label{model_df_nT_compare}
\end{figure}

The XENONnT experiment has not published information about their running parameters yet. Studies of the expected sensitivity and performance reported in ref.~\cite{XENON:2020kmp} are used to model the electronic recoil response within this framework. These predictions also rely on the efficiencies and energy resolution from the SR1 of XENON1T. These assumptions are expected to change when XENONnT reports their first results. It is expected that XENONnT will be an improvement from its predecessor so that this framework is therefore following the conservative approach of ref.~\cite{XENON:2020kmp}. The expected live time of the first science run of XENONnT is not known. Several possible total live times starting from the approximate end of commissioning of the XENONnT detector in June 2021 are assumed: $1$ year, $3$ years and $5$ years. The XENONnT experiment will also need to characterize their detector response using calibrations similar to XENON1T so that the effective live time of background data taking can be assumed to be similar to XENON1T which amounts to a fraction of 0.61. An upper bound on the sensitivity estimates can be determined by assuming an effective live time fraction of $1.0$ which corresponds to background data taking only with no calibrations during the science run of XENONnT. The live time fraction scales the exposure of the detector and would correspond to a longer overall measurement duration with calibration time included. The targeted FV of XENONnT is 4000 kg. Changes in the actual FV will scale the exposure and have a similar impact on the estimations as shown by the effective live time fractions, assuming that the electronic recoil backgrounds in each FV are comparable. The resulting expectation from the theoretical annual modulation model from section~\ref{sec3} for the XENONnT experiment over three years with an effective live time fraction of $0.61$ is shown in Fig.~\ref{model_df_nT_compare}.

The electronic recoil background components of the XENON1T and XENONnT experiments are modeled using the same considerations as the theoretical annual modulation model. The framework used here has the ability to model each background according to its energy spectrum and time dependence. Due to the lack of information on the time dependence for each background component in XENON1T and using predictions of XENONnT only, combined background expectations are considered in this study. The combined expectations summed over all components are modeled as stable in time although the XENON1T experiment observed time-dependent backgrounds such as $^{83\textrm{m}}\textrm{Kr}$, $^{133}\textrm{Xe}$, $^{131\textrm{m}}\textrm{Xe}$ and $^{125}\textrm{I}$. This study conservatively assumes that the time-dependent backgrounds can be fully modeled and do not impact the detection of signal events so that backgrounds which are stable in time will result in a similar sensitivity to signal events. The exact knowledge of the time-dependent backgrounds would be needed for time and energy fits of the XENON1T excess data if XENON1T would publish the timestamps of all electronic recoil events. In addition, the energy range of interest is reduced to 0$-$30 keV in this framework in order to minimize the impact of other background sources and focus on the signal region which is expected to be below 5 keV depending on the choice of signal model parameters.

The background models for the XENON1T experiment are taken from ref.~\cite{Aprile:2020tmw}. This work focuses on using the combined background model called $B_0$~\cite{Aprile:2020tmwdata} which is interpolated and scaled over the energy range of $0$ to $30$ keV. The resulting estimated event rate from the model implementation is $66.32$ detected events/(tonne $\times$ year $\times$ keV) or $74.5$ events/(tonne $\times$ year $\times$ keV) after efficiency correction (meaning that all efficiencies are removed) while ref.~\cite{Aprile:2020tmw} reports $76 \pm 2$ events/(tonne $\times$ year $\times$ keV) for the same region of interest. The XENONnT background model was built using the background spectra of ref.~\cite{XENON:2020kmp} for each component. The given spectra were converted and/or calculated from their theoretical predictions and smeared using the given detector resolution yielding an average background level of about $12$ events/(tonne $\times$ year $\times$) keV at low electron recoil energies. 

\begin{figure}[H]
\centering
\includegraphics[width=0.69\textwidth]{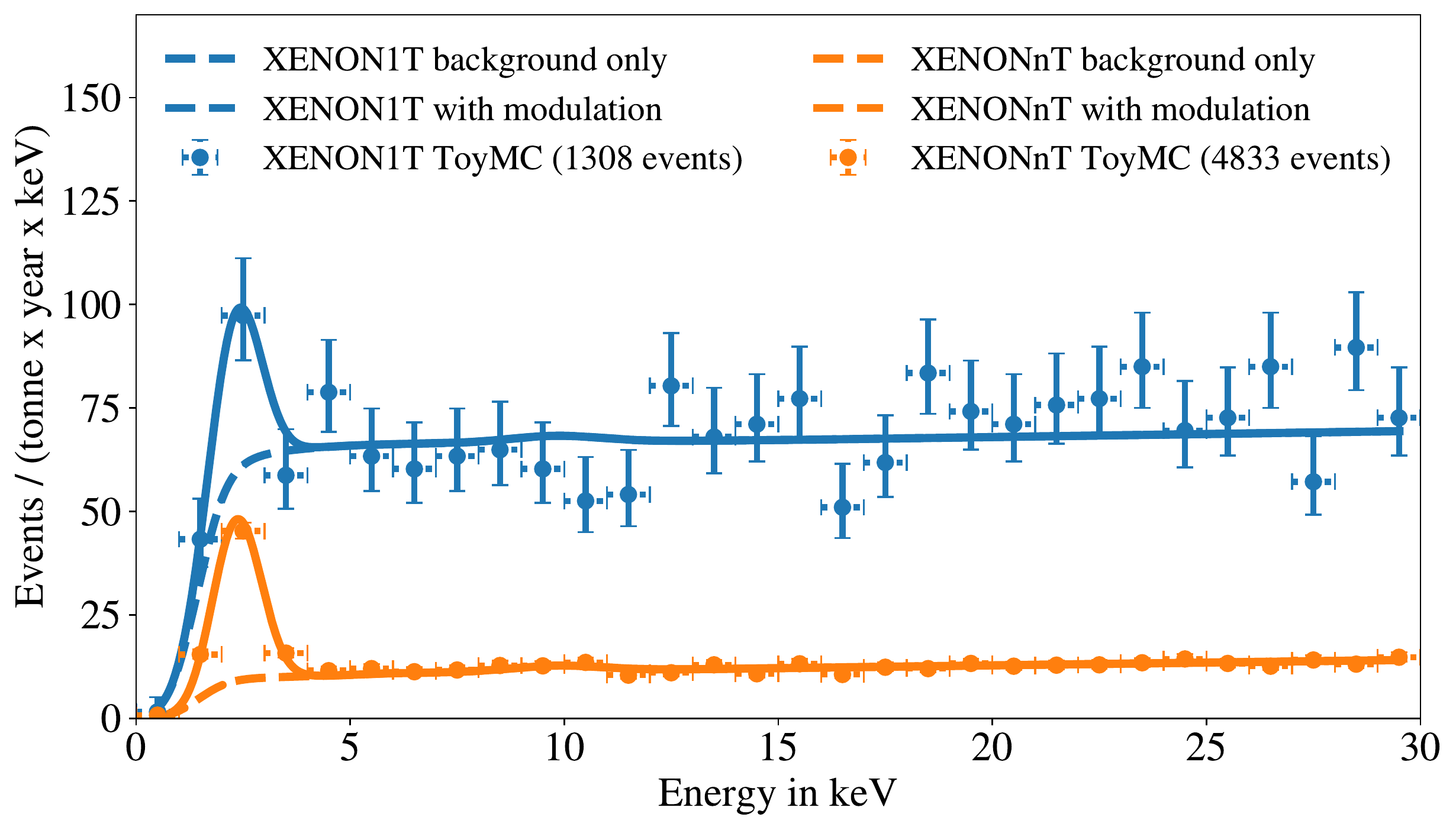}
\caption{Comparison of the annual modulation signal and combined background model predictions for the XENON1T and XENONnT experiments assuming $m_2 = 0.3\,\textrm{GeV}$, $\Delta m = 2.23\,\textrm{keV}$ and $\bar{\sigma}_e = 3.36\times10^{-44}$ cm$^2$ for the theoretical model. The solid line indicates the total observed event rate of the combination of signal and background events while the dashed line shows the background-only expectation. The data points illustrate the result of a single ToyMC (toy Monte Carlo) as described in section~\ref{sec6}.}
\label{bkg_sig}
\end{figure}

The annual modulation signal and combined background model predictions used in this analysis are given in Fig.~\ref{bkg_sig} for a specific set of theoretical model parameters. The rate of detected background events/(tonne $\times$ year $\times$ keV) for XENONnT is predicted to be about an order of magnitude smaller than XENON1T while the annual modulation signal strength is the same in both experiments due to the assumed detector response. 

The background and theoretical annual modulation models for the XENON1T experiment implemented in this framework can be used to describe the excess of electronic recoil events as reported in ref.~\cite{Aprile:2020tmw}. The electronic recoil event list from ref.~\cite{Aprile:2020tmwdata} only contains information about the energy of each event that occurred during the SR1 of XENON1T within a certain energy range but does not report the specific timestamp of an event. Thus, fitting the excess in energy and time is not possible. One can still exploit the energy space and obtain potential parameters of the theoretical annual modulation model which would fit the excess energy spectrum. This approach neglects the additional sensitivity from the time-dependent features of the modulation model presented in this work but allows one to reduce the parameter space. In sections \ref{sec2} and \ref{sec6} we highlight the time-dependent modulation features in event rate and energy. Further reduction of the parameter space would be possible by exploiting also the time information. 

\begin{figure}[H]
\centering
\includegraphics[width=0.99\textwidth]{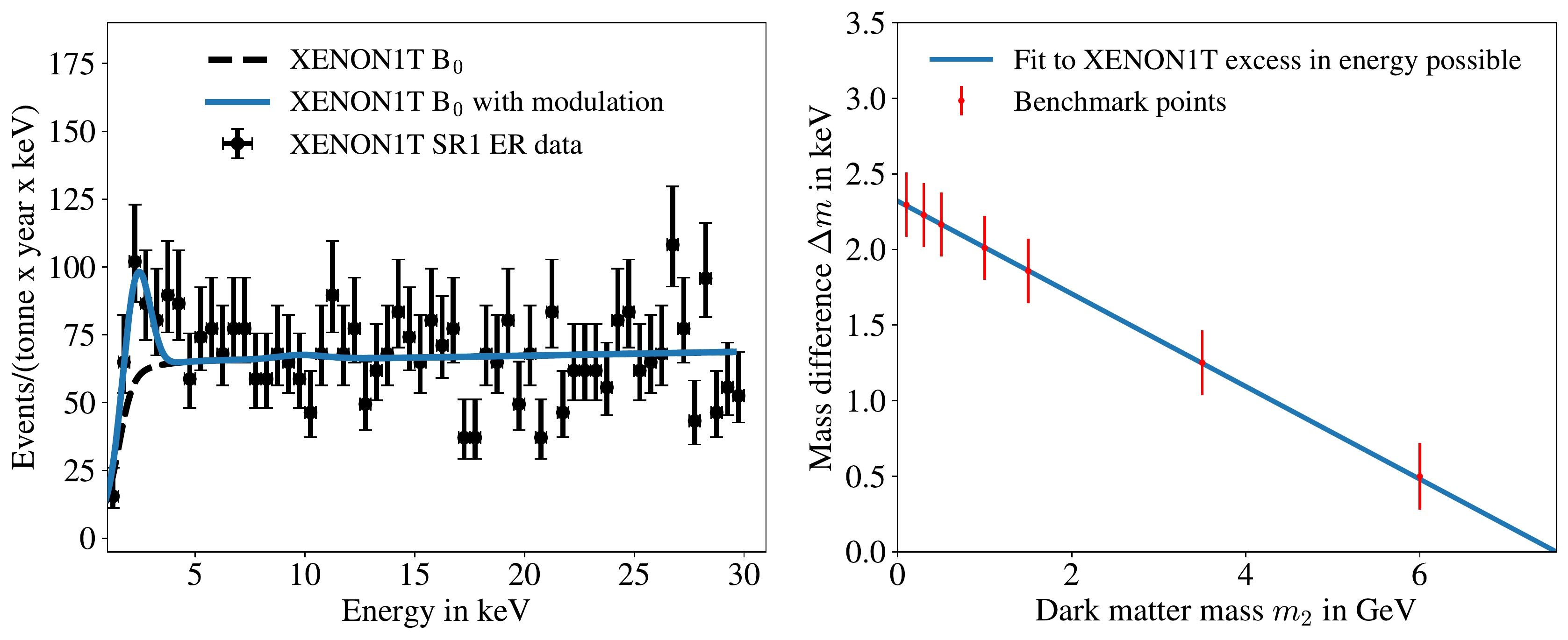}
\caption{Left panel: Illustration of a single energy-only extended unbinned likelihood fit to the XENON1T SR1 electronic recoil data~\cite{Aprile:2020tmw} using this framework. All fitted curves using the parameters reported in table~\ref{table_fit_res} are on top of each other and depicted by the blue line. The data points with $0.5\,\textrm{keV}$ binning are given as illustration of the event rate and are not used in the fitting. Right panel: Range of possible $m_2$ and $\Delta m$ combinations that can fit the excess as reported in ref.~\cite{Aprile:2020tmw}.}
\label{fit_sig_bkg}
\end{figure}

The models described earlier are summed over the time dimension in order to perform an unbinned extended likelihood fit of the electronic recoil energy spectrum from SR1 of XENON1T. To perform the fit, we follow the same approach as presented in section \ref{sec6}. Seven different dark matter masses $m_2$ are selected as benchmark points and the free parameters $\Delta m$ and $\bar{\sigma}_e$ are fitted. The results are reported in table~\ref{table_fit_res} and Fig.~\ref{fit_sig_bkg}. As seen in section \ref{sec2}, higher values of the dark matter mass $m_2$ will shift the peak position to higher recoil energies which can be compensated for by adjusting the mass difference $\Delta m$ to smaller values. This dependence is given by
\begin{align} \label{eq:1T_fit_energy}
\Delta m &= E_{\textrm{Excess}} - \frac{1}{2} m_2 (v_\textrm{ES}^2 + v_\textrm{SH}^2),
\end{align}
where $E_{\textrm{Excess}} = 2.32\,\textrm{keV}$ is the maximum electron recoil energy. Therefore one can find for each dark matter mass $m_2$ a corresponding mass difference $\Delta m$ value which results in a mean energy peak position in agreement with the electronic recoil excess reported by XENON1T. The dark matter electron scattering cross section $\bar{\sigma}_e$ enters the $\frac{dR}{dE\,dt}(E,t)$ calculation as a scaling factor and compensates for the change in the dark matter mass $m_2$.

This approach of only using the energy space is not sensitive to the energy modulation given in Eqs.~(\ref{relv}) and~(\ref{ER}). The modulation of the electronic recoil energy would result in a broadening of the energy spectrum much smaller than the detector resolution at low energies. However, access to the time information of the electronic recoils would allow to further characterize the dark matter mass $m_2$ space.

\begin{table}[H]
\centering
\begin{tabular}{|c|c|c|c|}
\hline
Model & $m_2$ [GeV] & $\Delta m$ [keV] & $\bar{\sigma}_e$ [cm$^2$] \\
\hline
I & $0.1$ & $2.30$ & $1.09\times 10^{-44}$ \\
\hline
II & $0.3$ & $2.23$ & $3.36\times 10^{-44}$ \\
\hline
III & $0.5$ & $2.16$ & $5.66\times 10^{-44}$ \\
\hline
IV & $1.0$ & $2.01$ & $1.14\times 10^{-43}$ \\
\hline
V & $1.5$ & $1.86$ & $1.71\times 10^{-43}$ \\
\hline
VI & $3.5$ & $1.25$ & $4.03\times 10^{-43}$ \\
\hline
VII & $6.0$ & $0.50$ & $7.07\times 10^{-43}$ \\
\hline
\end{tabular}
\caption{Results for energy-only fits of the XENON1T SR1 electronic recoil data~\cite{Aprile:2020tmw} using this framework. The dark matter mass $m_2$ is fixed and $\Delta m$ and $\bar{\sigma}_e$ are fitted using the theoretical annual modulation model.}
\label{table_fit_res}
\end{table}

\section{Simplified annual modulation model\label{sec5}}

The calculation of the fully featured theoretical annual modulation model described in section \ref{sec2} is computationally expensive and the provided level of precision is not needed for most experiments. Here, a simplified annual modulations model is proposed which incorporates the important time and energy signatures which drive the sensitivity. The aim of the simplified model is to test if a certain annual modulation signature in energy or time is present in the experimental data which would indicate the usage of the full theoretical model. 

Therefore, the simplified model only describes the shape of the signal and all quantities that scale the overall detected event rate $\frac{dR}{dE\,dt}(E,t)$ are combined in a single scaling parameter $\mathcal{N}$. The electronic recoil energy spectrum defined by $\eta(E_v,t) \, K'(E_v, t)$ in Eq.~(\ref{dRdE}) exhibits a narrow peak with a width of $\sim\,0.1\,\textrm{keV}$ and will be modeled by a mono-energetic energy deposition. The width of the theoretical energy spectrum is much smaller than the typical detector resolution $\sigma_E(E)$ at low energies for liquid xenon dual phase time projection chambers. As a consequence, the smeared energy spectra of the full and simplified theoretical models are comparable. The simplified model is not applicable at the detection threshold where detector efficiencies would alter the Gaussian shape of the full theoretical energy spectrum and thus impact the overall scaling.

The proposed simplified model can be written as
\begin{align}
\label{simplified1}
\frac{d\mathcal{R}}{dE\,dt}(E,t) &= \mathcal{N} \, \mathcal{R}_S(E, t) \, [1 - 2A \cos(2 \pi t - \phi)], \\
\mathcal{R}_S(E, t) &= \frac{1}{\sqrt{2\pi}\sigma_E(E_p(t))}  \exp\left[-\frac{(E-E_p(t))^2}{2 \sigma^2_E(E_p(t))}\right] \alpha(E), \\
E_p(t) &= E_0 \, [1 + E_A \cos(2 \pi t - \phi)],
\label{simplified2}
\end{align}
where the scaling parameter $\mathcal{N}$ corresponds to the number of events in 1 tonne $\times$ year exposure after efficiency correction, $\mathcal{R}_S(E, t)$ gives the energy spectrum incorporating the detector resolution and detector efficiencies, and $E_p(t)$ is the modulating recoil energy. The central electronic recoil energy of $\mathcal{R}_S(E, t)$ is given by $E_0$ and the modulation amplitude in energy by $E_A$. The fractional event rate modulation amplitude is given by $A$. The quantities $\phi$, $\sigma_E(E)$ and $\alpha(E)$ are the same as described in sections \ref{sec2} and \ref{sec4}. The detector resolution $\sigma_E(E)$ has to be treated as a free parameter if the width of the theoretical energy spectrum is of similar size at the energies of interest.

\begin{figure}[H]
\centering
\includegraphics[width=0.7\textwidth]{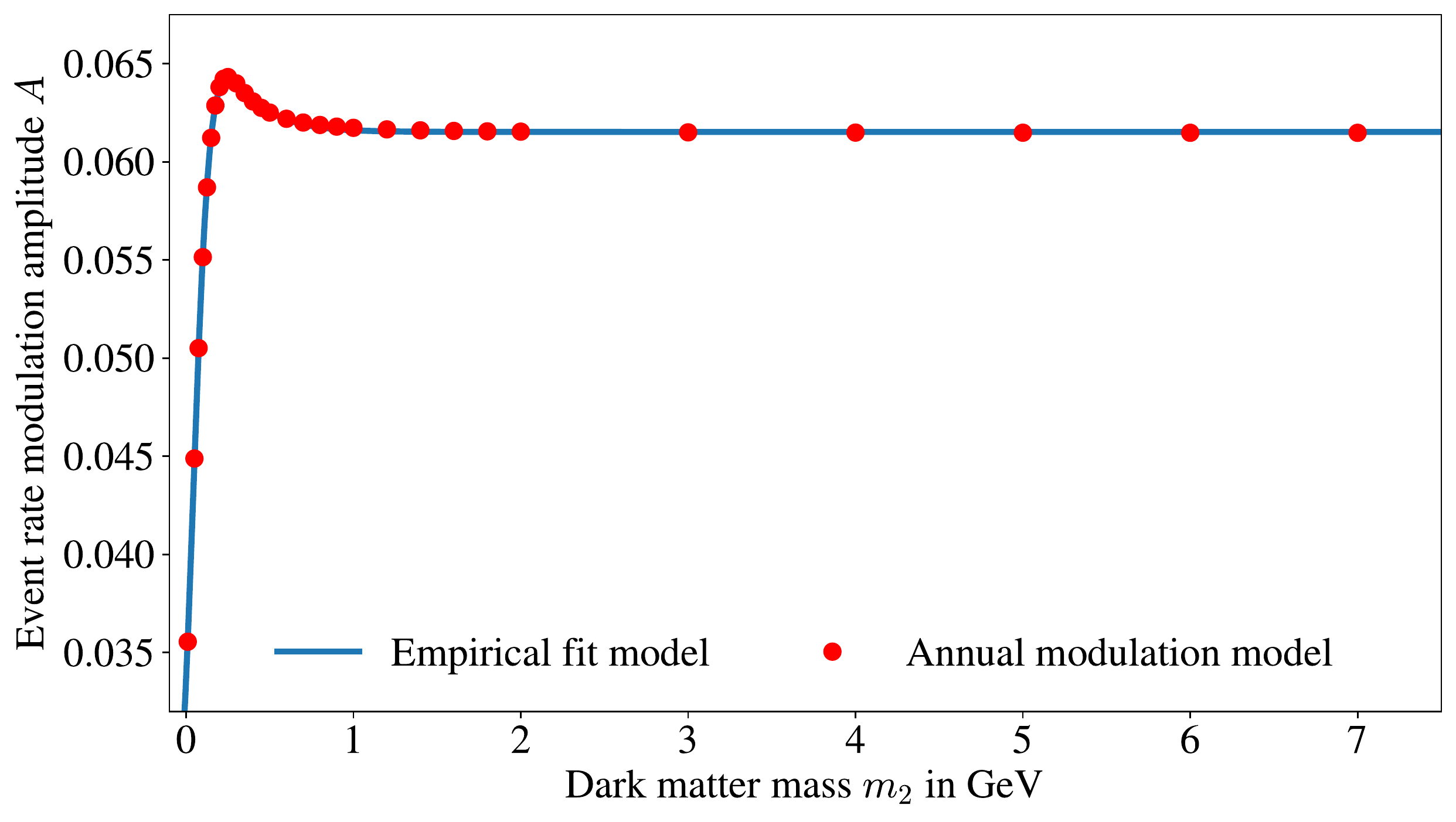}
\caption{Fractional modulation amplitude $A$ versus the dark matter mass $m_2$ calculated using the full theoretical annual modulation model as given in section \ref{sec2} (red dots). An empirical model (blue line) is used as a parameterization.}
\label{toymodulation_A_m2_fit}
\end{figure}

It is possible to directly link the physical parameters $\Delta m$ and $m_2$ to the shape parameters $E_0$, $E_A$ and $A$ of the simplified model in Eqs.~(\ref{simplified1}) and~(\ref{simplified2}). The central electronic recoil energy $E_0$ and the modulation amplitude in energy $E_A$ can be directly isolated in the formulae given in section \ref{sec2}, so that
\begin{align}
E_0 &= \Delta m + \frac{1}{2} m_2 (v_\textrm{ES}^2 + v_\textrm{SH}^2), \label{E_0} \\
E_A &= \frac{m_2 \, v_\textrm{ES} \, v_\textrm{SH}\,\cos\gamma}{E_0}. \label{E_A}
\end{align}
The fractional event rate modulation amplitude $A$ was found to be only dependent on $m_2$ in energy ranges where the simplified model is applicable. Therefore, the theoretical model was evaluated at benchmark points spanning the full possible range of dark matter masses $m_2$ as shown in Fig.~\ref{toymodulation_A_m2_fit}. An empirical model was used to parameterize the obtained dependence given by
\begin{align}
A &= \frac{y}{1 + e^{-z (m_2-k)}} + r \, e^{s \, m_2}, \label{A}
\end{align}
where $y=0.0615$, $z=19.0225/\textrm{GeV}$, $k=0.0373\,\textrm{GeV}$, $r=0.0133$ and $s=-5.0639/\textrm{GeV}$. 

The parameters $E_0$, $E_A$ of the simplified model illustrate the possibility of the full theoretical model to constrain the dark matter mass $m_2$ directly from the modulation in the electronic recoil energy spectrum. In addition, the empirical function for $A$ shows that the event rate modulation is mostly driven by $m_2$ which can be exploited to further increase the sensitivity to $m_2$ by using the full time and energy information of the electronic recoils in an experiment.

\section{Analysis methods and results}\label{sec6}

\subsection{Statistical method}

In this analysis, we use the unbinned profile likelihood method. The likelihood is constructed as
\begin{align} 
\mathcal{L}(\mu_b, \mu_s, \vec{\theta}) = &\text{\,Poiss}(N|(\mu_b+\mu_s)) \nonumber \\
&\times \prod_{i}^{N}\left(\frac{\mu_b}{(\mu_b+\mu_s)} \, f_b(\vec{p}_i) + \frac{\mu_s}{(\mu_b+\mu_s)} \, f_s(\vec{p}_i, \vec{\theta}) \right),
\label{eq:basic_likelihood}	
\end{align}
where $\mu_s$ and $\mu_b$ are the expected total number of signal and background events, the index $i$ runs over the total number of observed events in data $N$ and $f_b$ and $f_s$ are the probability density functions of the signal and background models. The nuisance parameters of the full model, $m_2$, $\Delta{m}$ and $\bar{\sigma}_e$, and the simplified model, $\mathcal{N}$, $E_0$, $E_A$ and $A$, are embedded in $\vec{\theta}$. The framework incorporates the possibility to utilize the signal and background models in a 2-dimensional or 1-dimensional space where $\vec{p}_i$ consists of the energy $E_i$ and/or timestamp $t_i$ of the $i$th event.

\begin{figure}[H]
\centering
\includegraphics[width=0.48\textwidth]{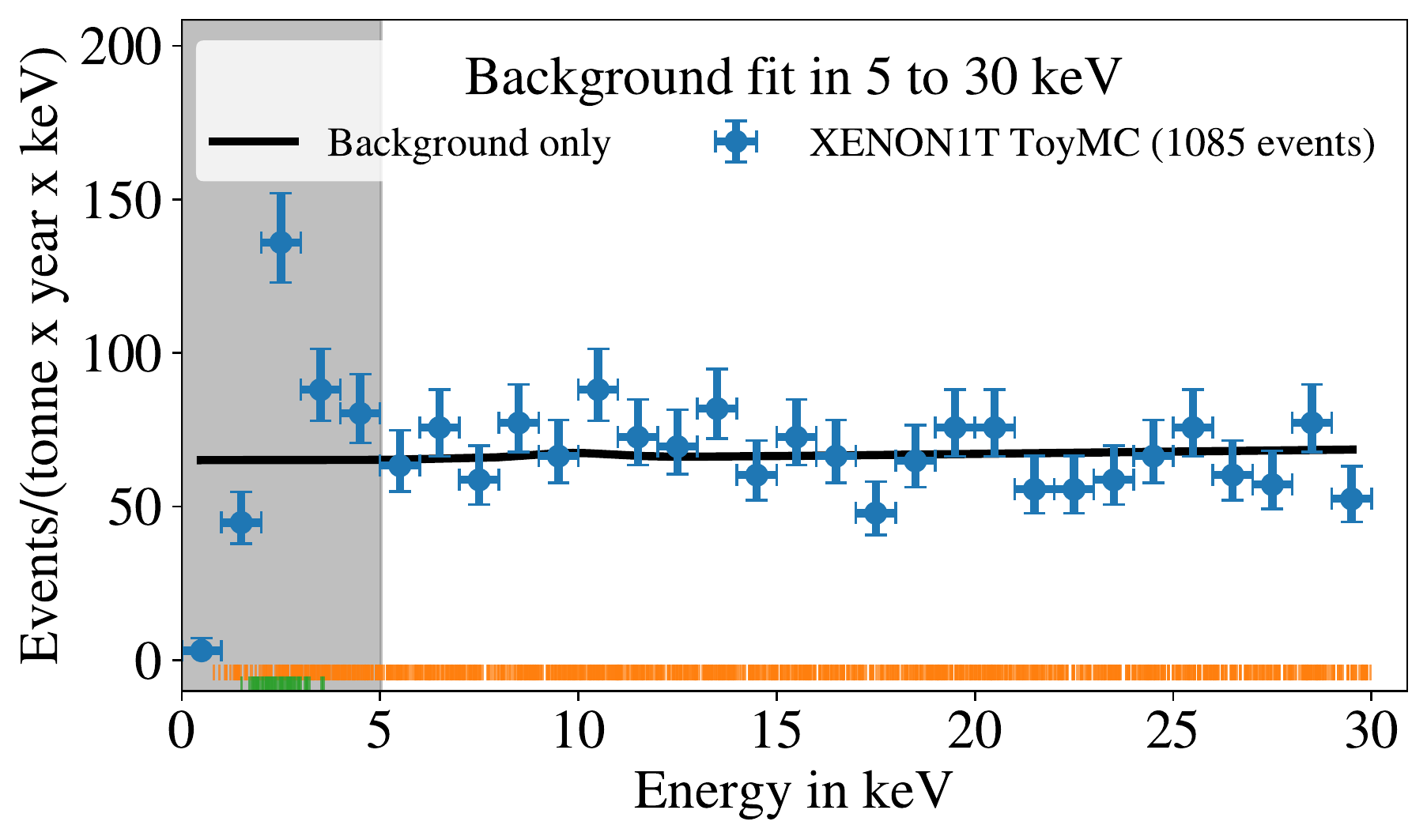}
\includegraphics[width=0.49\textwidth]{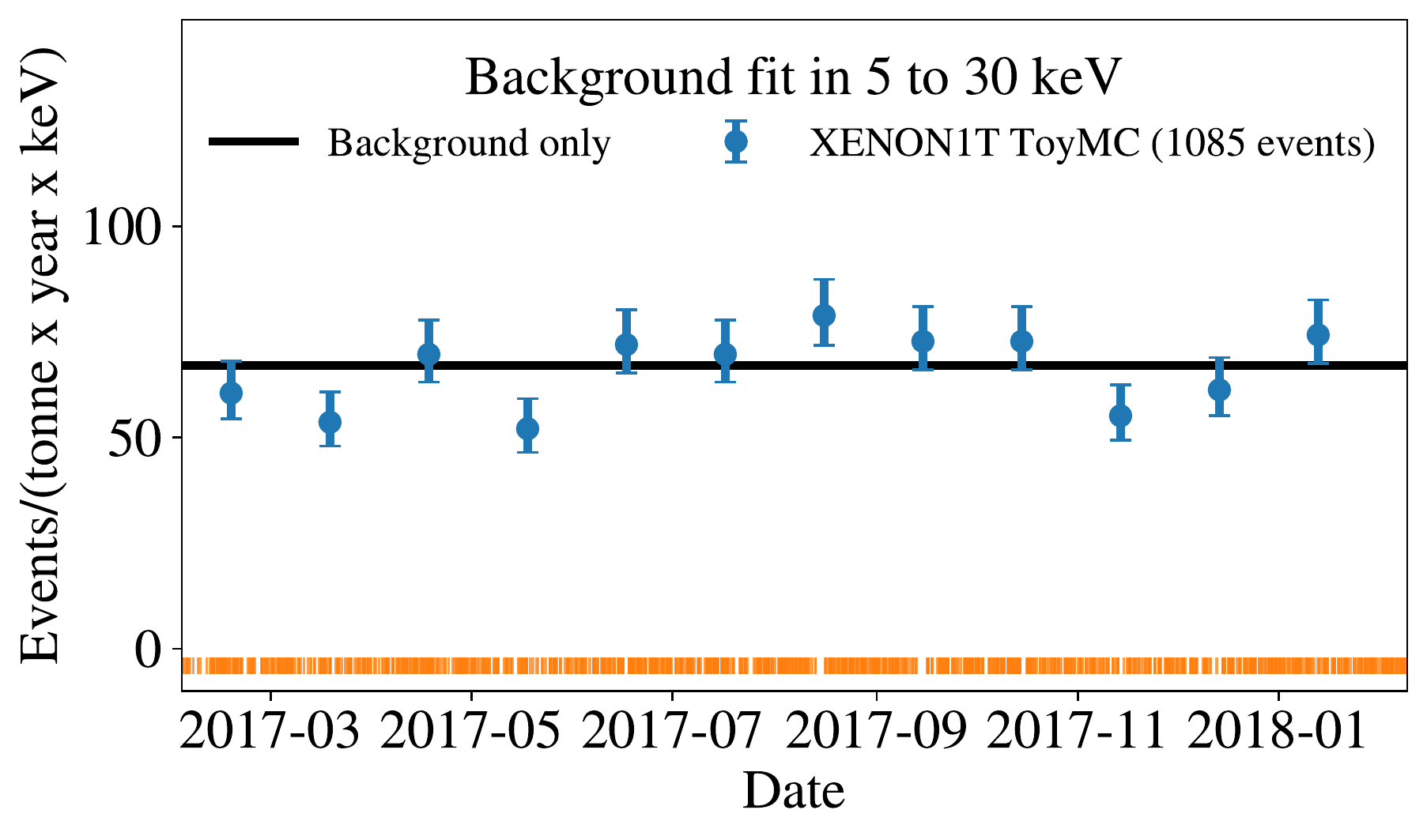}
\includegraphics[width=0.49\textwidth]{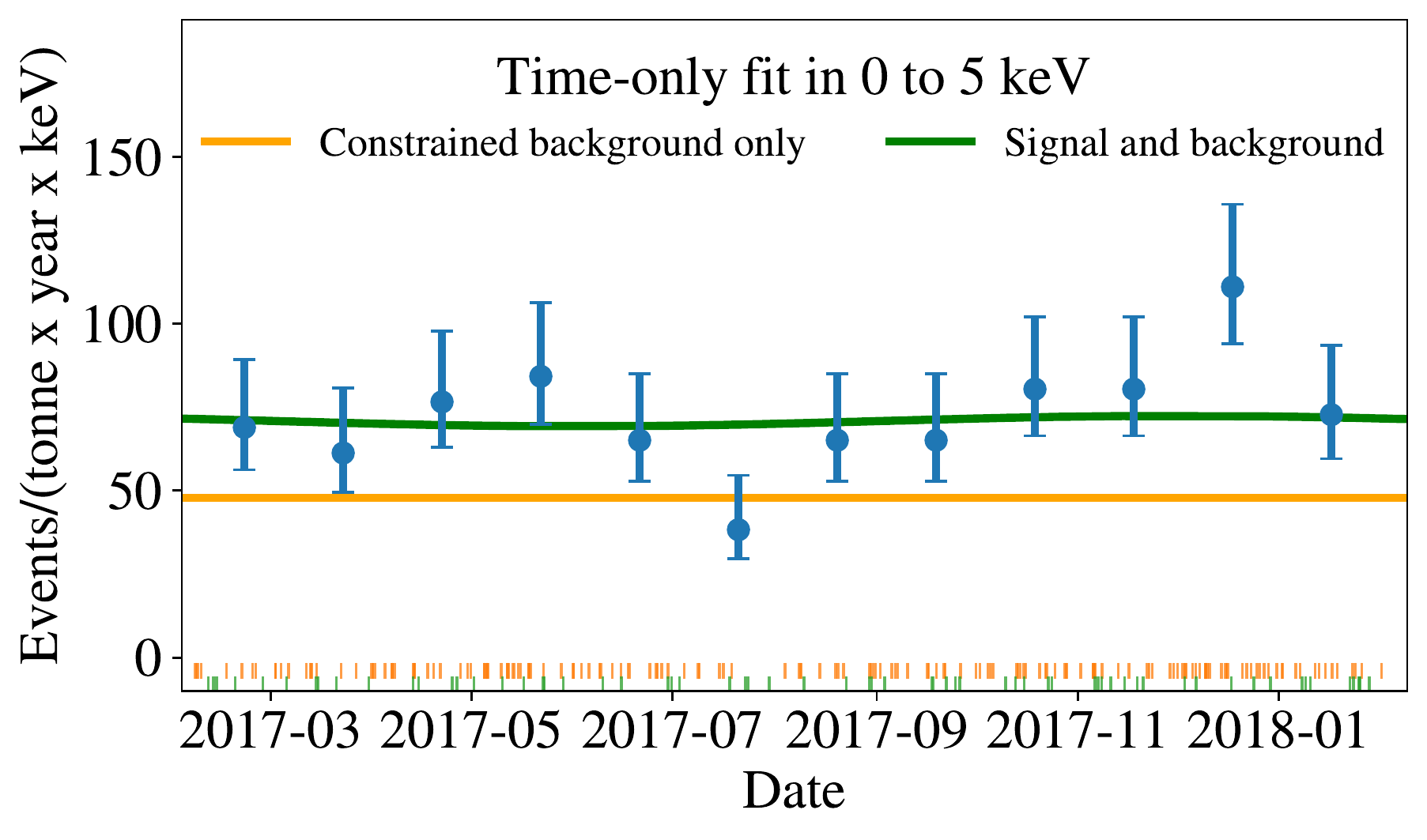}
\caption{Illustration of a time-only fit  using $\mathcal{L}_E(\mu_{b_E})$ (upper row) and $\mathcal{L}_t(\mu_b, \mu_s, \vec{\theta})$ (lower row) for simulated ToyMC data similar to XENON1T drawn from $f_b$ and $f_s$ over the full energy range of 0 to 30 keV. The binned event rates (in blue) are only for illustration purposes since the analysis is carried out using an unbinned likelihood method. Event distributions for background (orange) and signal (green) events are indicated by vertical lines.}
\label{sens_time_only}
\end{figure}

Searches for annual modulations are often carried out in time space only \cite{Bernabei:2018jrt,XENON100:2015hgt}. Annual modulation models usually consist of a sinusoidal modulation on top of a constant offset for the detected signal event rate. These modulation searches combine the parts of the signal and background components which are constant in time and perform a free fit to the sum. The knowledge of the background components from the full energy spectrum which extend into the signal region are neglected. This methodology reduces the sensitivity for modulation signals to only the fraction of the event rate which shows modulations in time over the signal energy range. Time-only studies within this framework will make use of the background information from the full available energy space and thus profit from an increased sensitivity by also including the constant signal offset in the modelling. A two-step approach is utilized where the background models in the time-only likelihood are constrained by the energy range outside of the signal region. The likelihoods for the time-only studies are constructed as
\begin{align} 
\mathcal{L}_E(\mu_{b_E}) = &\text{\,Poiss}(N|(\mu_{b_E})) \times \prod_{i}^{N} \, f_b(E_i), \label{eq:time_E} \\
\mathcal{L}_t(\mu_b, \mu_s, \vec{\theta}) = &\text{\,Poiss}(N|(\mu_b+\mu_s)) \nonumber \\
&\times \prod_{i}^{N}\left(\frac{\mu_b}{(\mu_b+\mu_s)} \, f_b(t_i) + \frac{\mu_s}{(\mu_b+\mu_s)} \, f_s(t_i, \vec{\theta}) \right) \label{eq:time_t} \\
&\times \, C_{\mu_{b_E}}, \nonumber
\end{align}
where $\mu_{b_E}$ is number of expected background events and $C_{\mu_{b_E}}$ the corresponding constraint. The likelihood $\mathcal{L}_E(\mu_{b_E})$ is applied in the background region from 5 to 30 keV and $\mathcal{L}_t(\mu_b, \mu_s, \vec{\theta})$ in the signal energy range from 0 to 5 keV. This methodology is illustrated in Fig.~\ref{sens_time_only} using simulated data (ToyMCs) drawn from the background and signal distribution, $f_b$ and $f_s$, for a specific amount of expected events and set of nuisance parameters. The background expectation $\mu_{b_E}$ from the background energy region is used to constrain the number of background events in the signal region. This approach assumes that all backgrounds can be sufficiently described outside the signal energy range. It is shown that a constrained background-only fit in the signal energy range underestimates the number of detected events compared to $\mathcal{L}_t(\mu_b, \mu_s, \vec{\theta})$. This method has the advantage of utilizing the increase in constant event rates but neglects the time-dependent energy modulations which are only accessible using $\mathcal{L}(\mu_b, \mu_s, \vec{\theta})$ from Eq.~(\ref{eq:basic_likelihood}) with $\vec{p}_i = (E_i, t_i)$.

Studies carried out in the energy and time space are performed using Eq.~(\ref{eq:basic_likelihood}) where $f_s$ corresponds to specific annual modulation model given by $\frac{dR}{dE\,dt}(E,t)$ or $\frac{d\mathcal{R}}{dE\,dt}(E,t)$. Time-only estimates are derived with respect to Eqs.~(\ref{eq:time_E}) and~(\ref{eq:time_t}) by integrating the signal and background models over the time and/or energy space.

\subsection{Sensitivity studies}\label{sec6.2}

The sensitivities of the XENON1T and XENONnT experiments are obtained by likelihood ratio tests using the test statistic
\begin{align}
q(\mu_s) &= -2 \ln\left(\frac{\mathcal{L}(\hat{\hat{\mu_b}}, \mu_s, \hat{\hat{\vec{\theta}}})}{\mathcal{L}(\hat{\mu_b}, \hat{\mu_s}, \hat{\vec{\theta}})}\right), 
\end{align}
where quantities with a single hat denote the set of parameters which correspond to the unconditional maximum of the likelihood while double-hatted quantities denote the set of parameters maximizing the conditional likelihood. Parameters without a hat are fixed to their true value. 

Under certain conditions the test statistic, $q(\mu_s)$, follows an asymptotic distribution which is given by a $\chi^2$ distribution with one degree of freedom~\cite{Cowan:2010js}. The distribution of $q(\mu_s)$ is estimated using ToyMCs drawn from the background and signal distribution, $f_b$ and $f_s$, to validate the assumption of asymptoticity. The sensitivities of the XENON experiments are derived as $90\,\%$ CL upper limits by performing likelihood scans over the range of dark matter masses $m_2$ and mass differences $\Delta m$ for various combinations of live times. The limit is obtained by the intersection of the likelihood scans with the 90\% quantile of a $\chi^2$ with one degree of freedom at $q(\mu_s)=2.71$.

\begin{figure}[H]
\centering
\includegraphics[width=0.85\textwidth]{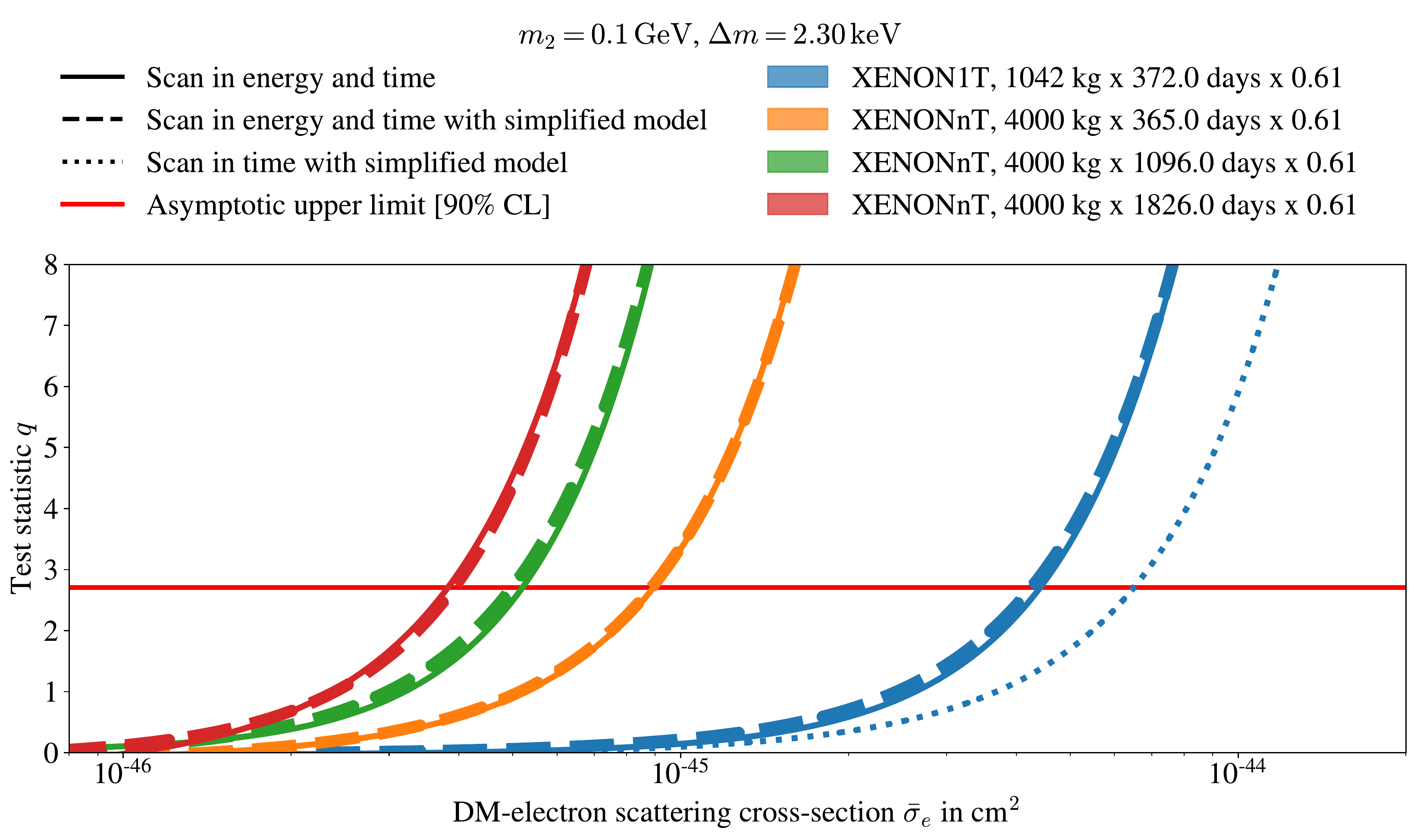}
\caption{Results of the profile likelihood scans over $\bar{\sigma}_e$ for $m_2 = 0.1\,\textrm{GeV}$ and $\Delta m = 2.30\,\textrm{keV}$ for various running conditions of the XENON1T and XENONnT experiments. Scans with the full theoretical model (solid), the simplified model (dashed) and the 1.5 times less sensitive time-only (dotted) are evaluated. The $90\,\%$ CL upper limit on $\bar{\sigma}_e$ at $q(\mu_s)=2.71$ is indicated by the red line.}
\label{likelihood_scan}
\end{figure}

As described in previous sections, the observed rate and energy modulation shape is driven by the parameter $m_2$ and the number of detected signal events driven by $\bar{\sigma}_e$. The parameter $\Delta m$ is only shifting the energy spectrum and does not impact the sensitivity if the signal is observed above the detection threshold of the experiment. The results of the profile likelihood scan for the selected benchmarks are given in Fig.~\ref{likelihood_scan}. The true values for $m_2$ and $\Delta m$ are selected to describe the XENON1T excess in electronic recoil events given in section \ref{sec5} and converted into the corresponding parameters of the simplified modulation model using Eqs.~(\ref{E_0}),~(\ref{E_A}) and~(\ref{A}). The number of signal events $\mathcal{N}$ of the simplified model is inferred from the full theoretical model using the truth value of $\bar{\sigma}_e$ in order to directly compare the sensitivity of both models. The number of detected signal events is mostly driven by $\bar{\sigma}_e$. Each likelihood scan over $\bar{\sigma}_e$ consists of $30$ points sampled in equal distance on a logarithmic scale. Each point corresponds to the result of $q(\mu_s)$ obtained from $1000$ ToyMCs from which the mean likelihood ratio and $1\sigma$ range are calculated. The results for XENON1T and XENONnT are given for effective live time fractions of $0.61$. The $90\,\%$ CL upper limit is indicated by the red line at $q(\mu_s)=2.71$. Scans with the full theoretical model show the same sensitivity as the simplified model so that the results are not differentiated in the following. As a comparison, a time-only scan is added for the XENON1T case in which the background was constrained following the methodology described above. The simplified model was used for this comparison and similar results can be achieved with the full theoretical model. 

\begin{figure}[H]
\centering
\includegraphics[width=0.85\textwidth]{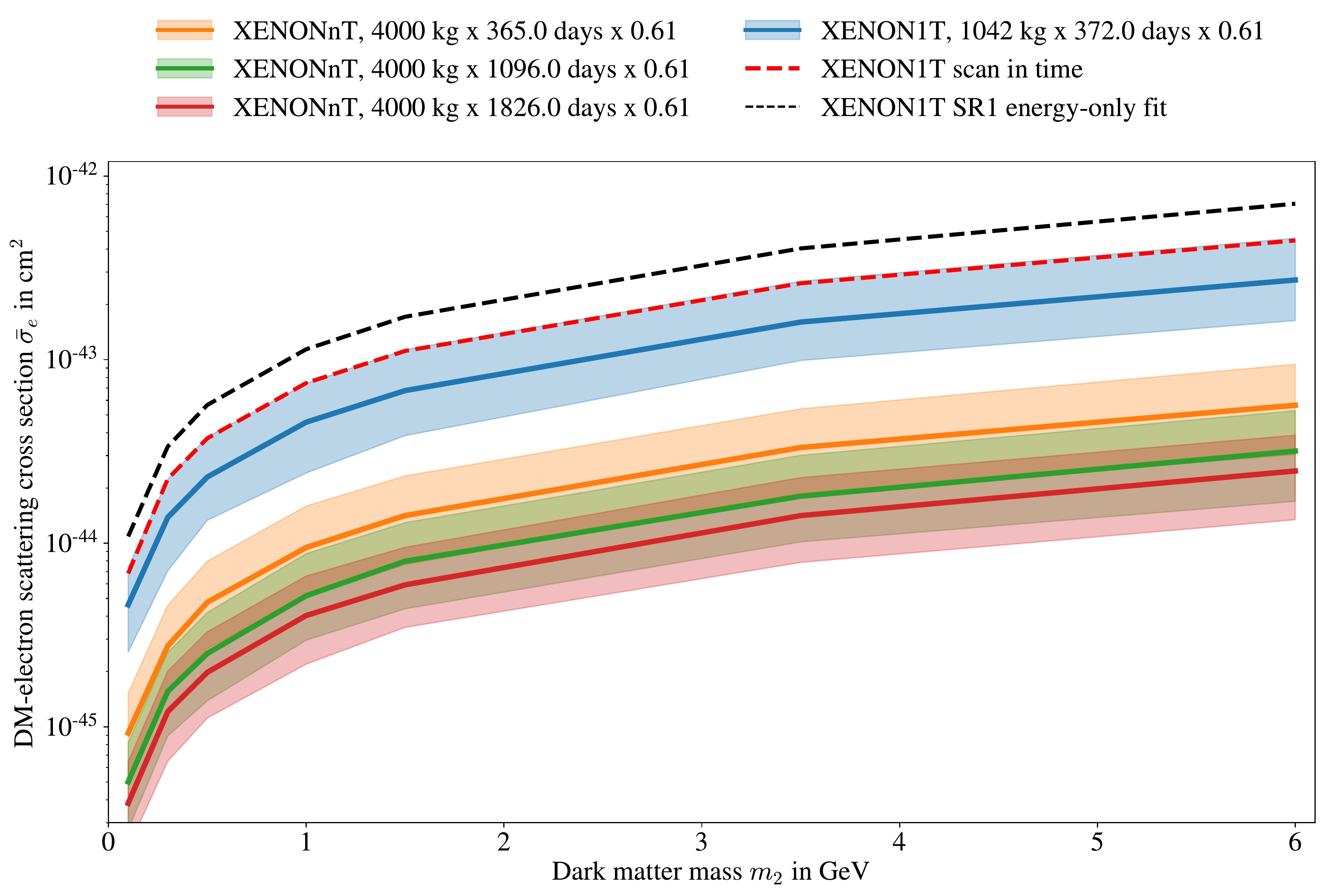}
\caption{Result for the $90\,\%$ CL upper limit on $\bar{\sigma}_e$ versus the dark matter mass $m_2$ for ToyMC data for various running conditions of the XENON1T and XENONnT experiments. Scans with the full theoretical or simplified model (solid) are evaluated and their $1\sigma$ statistical uncertainty on the $90\,\%$ CL upper limits are given as shaded regions. The energy-only best fit curve as described in section \ref{sec5} and the result for the about 1.5 times less sensitive time-only scan are given as reference.}
\label{sens_combined}
\end{figure}

Limits derived from the time-only method are consistently by a factor 1.5 times lower than the full two-dimensional models indicating that the knowledge of the energy spectrum as well as its time modulations will boost the achieved sensitivities. This translates to a smaller exposure as kg\,$\times$\,days of data to achieve the same level of sensitivity. XENONnT will already exhibit an improvement in sensitivity compared to the SR1 of XENON1T after a few months of live time depending on the specific date range due to the dependence on the maximum event rate for annual modulations. The sensitivity of XENONnT after 1 year of data taking and $222.65\,\textrm{days}$ of live time is a factor five lower than for the SR1 of XENON1T. The obtained results can be illustrated in the $m_2$ and $\bar{\sigma}_e$ space as given in Fig.~\ref{sens_combined}, assuming that $\Delta m$ has no influence on the limit setting as described above.

\begin{figure}[H]
\centering
\includegraphics[width=0.99\textwidth]{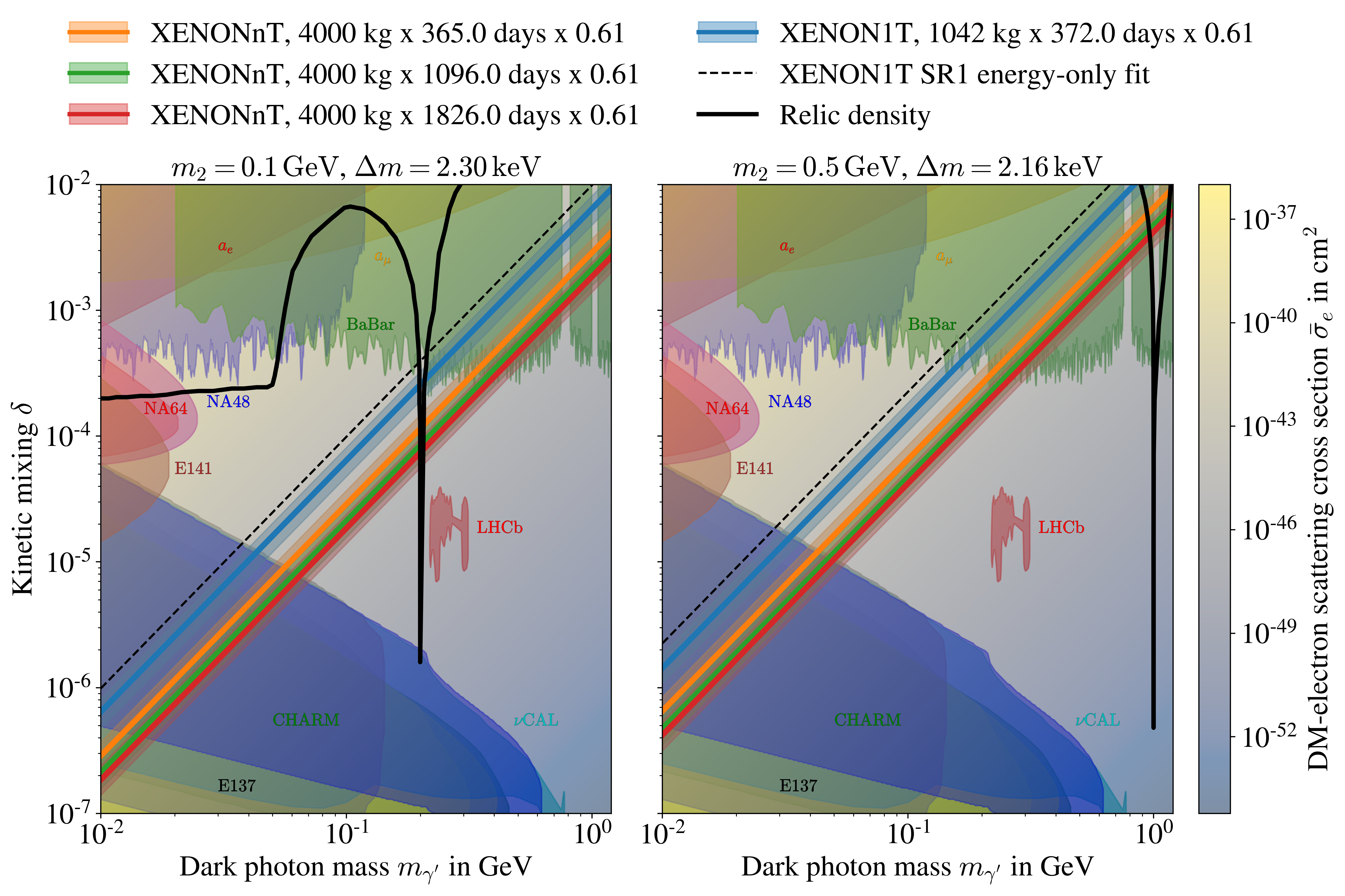}
\caption{Result for the $90\,\%$ CL upper limit on $\bar{\sigma}_e$ in the space of dark photon mass $m_{\gamma'}$ and kinetic mixing $\delta$. The shaded regions around the diagonal lines indicate the $1\sigma$ statistical uncertainty. Here we take $g_X=0.01$ and $m_2$ and $\Delta m$ values are indicated in the legends of each plot. Constraints on the dark photon from beam dump experiments and others are obtained from \code{darkcast}~\cite{Ilten:2018crw}. The solid black line corresponds to regions of the parameter space satisfying the DM relic density.}
\label{sens_combined_sigma_e}
\end{figure}

The 90\% CL fit curve from the energy-only fit as described in section \ref{sec5} and the result for the time-only scan are given as reference. The results for the $90\,\%$ CL upper limit on $\bar{\sigma}_e$ can also be converted into the space of dark photon mass $m_{\gamma'}$ and kinetic mixing $\delta$ using Eq.~(\ref{inelastic-cross}), given in Fig.~\ref{sens_combined_sigma_e}.

An annual modulation fit in energy and time of actual data from the XENON1T and XENONnT experiments would allow to reduce the dark matter mass $m_2$ parameter space due to direct dependence of energy and time modulation on $m_2$ as seen in Eqs.~(\ref{dRdE}),~(\ref{eq:time_E}) and~(\ref{eq:time_t}). The performance of such fits is evaluated using ToyMCs injecting an annual modulation signal similar to the one observed in XENON1T SR1 data fulfilling Eq.~(\ref{eq:1T_fit_energy}). The simulated events are freely fitted using Eq.~(\ref{eq:basic_likelihood}) and the best-fit values for the parameters $m_2$, $\Delta m$ and $\bar{\sigma}_e$ are obtained. The best-fit values of the dark matter mass $m_2$ parameter in the case of XENON1T SR1 simulations exhibits a broad distribution around the true value with a $1\sigma$ uncertainty of $5.65\,\textrm{GeV}$ which spans the entire investigated parameter space. One year of data from XENONnT would allow to consistently recover the true value of the ToyMCs in the fits and reduce the uncertainty to $2\,\textrm{GeV}$. It is possible to obtain the dark matter mass $m_2$ with an uncertainty of $1\,\textrm{GeV}$ over the planned XENONnT lifespan of five years. Simulations for $m_2$ masses below $0.3\,\textrm{GeV}$ exhibits an additional distribution of fits centered around $m_2 = 0$ indicating that further improvements of the statistical treatment for low $m_2$ masses are needed.

The two-dimensional fits as well as fits in the time-only space to actual data obtained from an experiment can only be performed if the periods of good data taking are published alongside with the event lists containing the energy and time information. In addition, the live time efficiency for each data period is needed which is comparable to the detection efficiency in the energy space.

\begin{figure}[H]
\centering
\includegraphics[width=0.49\textwidth]{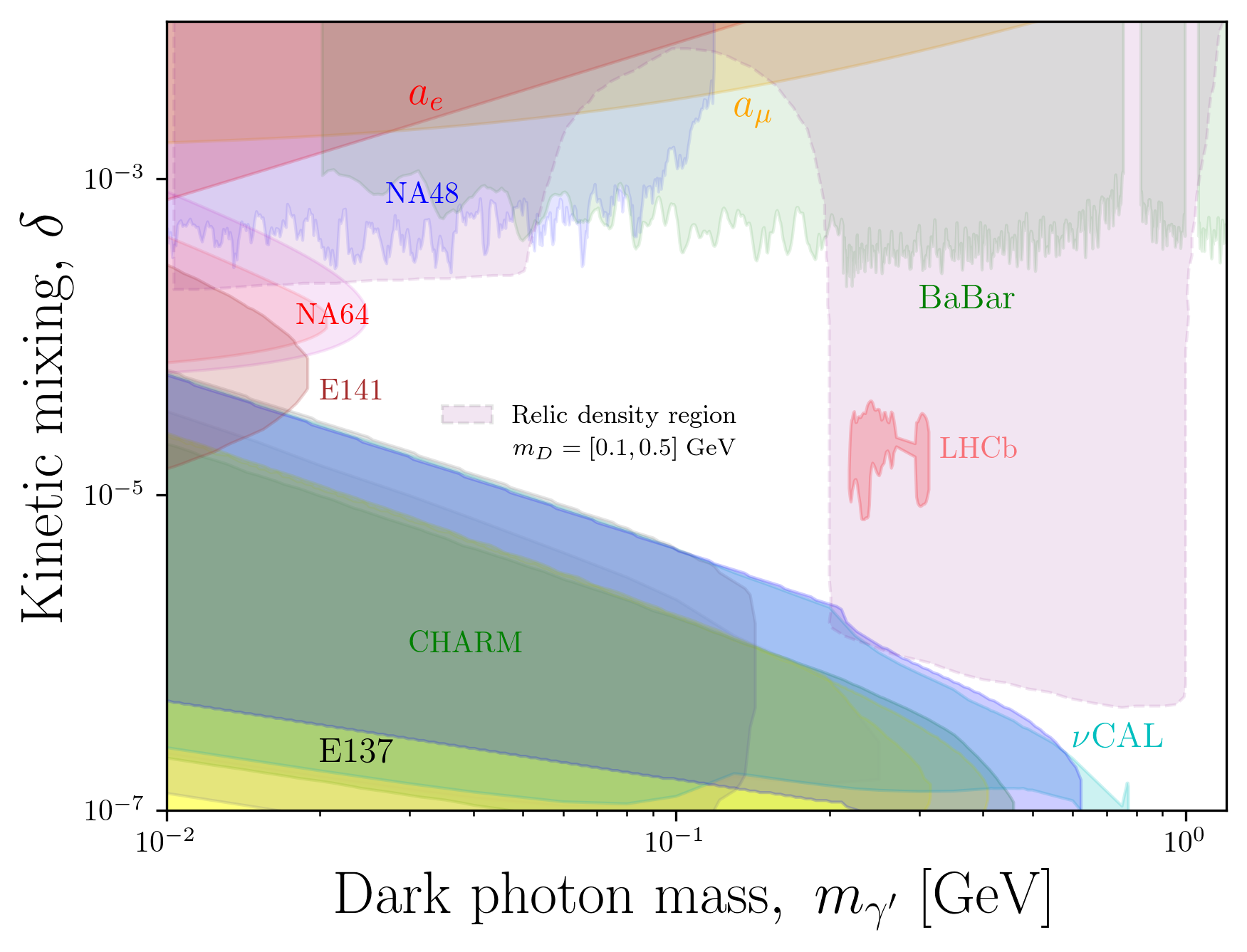}
\includegraphics[width=0.49\textwidth]{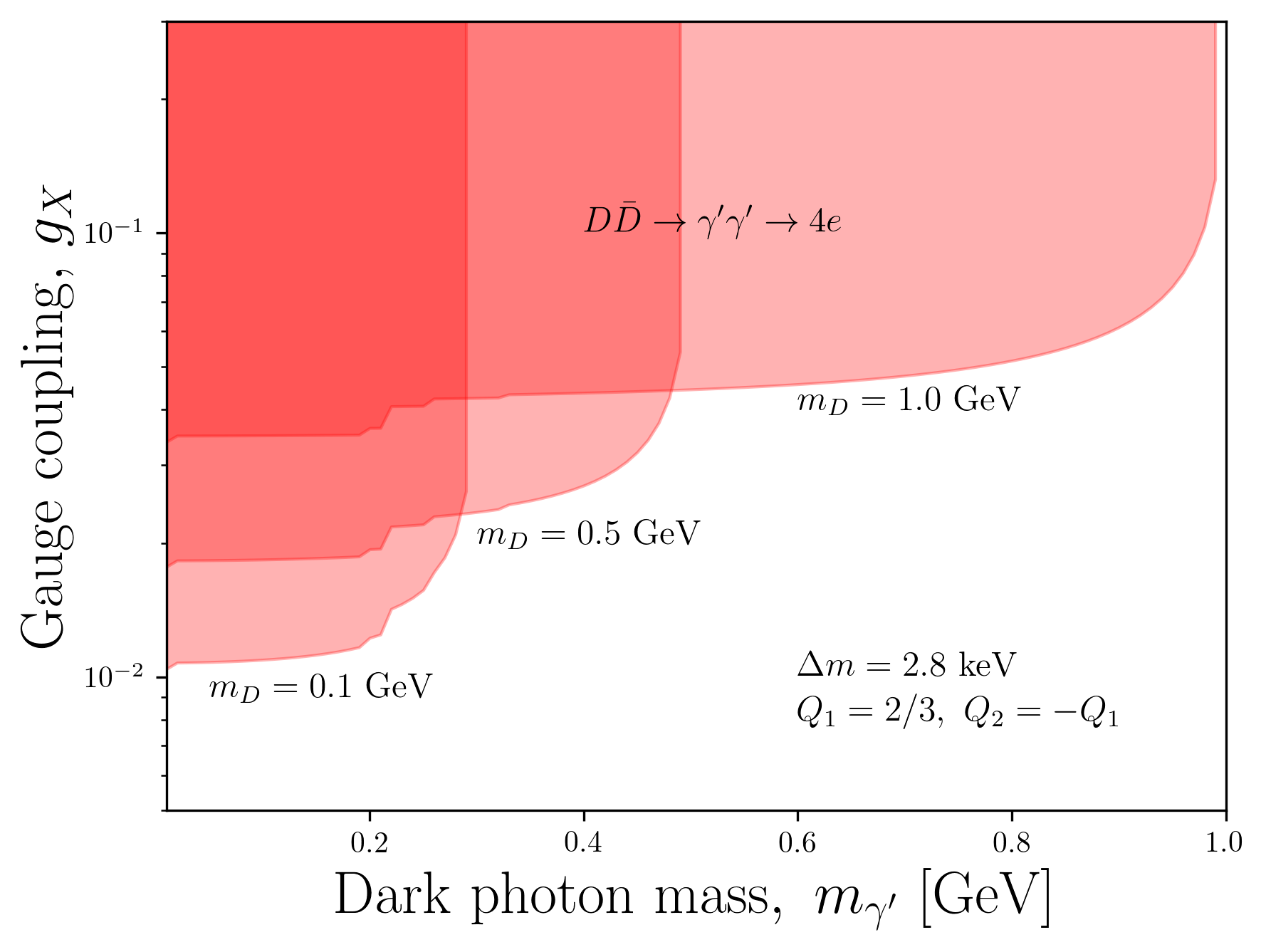}
\caption{Left panel: The region of allowed DM relic density for a range of DM masses and with $g_X=0.01$ in the kinetic mixing$-$dark photon mass plane. Other experimental constraints on the dark photon similar to Fig.~\ref{sens_combined_sigma_e} are shown. Right panel: Constraints on the dark gauge coupling from Planck recast to our model parameters in the DM annihilation to $4e$ channel. The limits are shown for three choices of DM mass.}
\label{kinetic_gX_limits}
\end{figure}

Let us now explain the other features of Fig.~\ref{sens_combined_sigma_e}. The black curves shown in both panels indicate the parts of the parameter space where the DM relic density is satisfied. For $m_2>m_{\gamma'}$, the DM relic density is mostly set by the annihilation process $D\bar{D}\to\gamma'\gamma'$ which is controlled only by $g_X$. Notice in the left panel of Fig.~\ref{sens_combined_sigma_e} how the black curve remains almost horizontal for small $m_{\gamma'}$ and the relic density barely changes since $g_X$ is fixed. The curve then experiences a sudden rise as $m_2$ becomes close to $m_{\gamma'}$. In this case, DM annihilation into SM particles, $D\bar{D}\to\gamma'\to f\bar{f}$, becomes relevant and a larger $\delta$ is required to deplete the relic density. The rise is followed by a sudden drop forming a funnel-like region in the vicinity of $m_{\gamma'}=2m_2$, i.e. near resonance. Since the DM annihilation cross section blows up, a smaller $\delta$ is required to maintain the correct relic density. As we move away from the resonance region, i.e. for $m_2<m_{\gamma'}$, larger $\delta$ values are again required to deplete the DM abundance via annihilation to the SM. In the right panel, the same funnel-like shape is seen close to $1\,\textrm{GeV}$ which is again indicative of DM depletion near resonance. Several parts of the parameter space satisfying the DM relic density are allowed by constraints on the dark photon from beam dump experiments and others. The constraints shown in Fig.~\ref{sens_combined_sigma_e} are derived from several experiments such as the electron and muon $g-2$~\cite{Endo:2012hp}, BaBar~\cite{Lees:2014xha}, CHARM~\cite{Bergsma:1985qz,Tsai:2019mtm,Gninenko:2012eq}, NA48~\cite{Batley:2015lha}, E137~\cite{Andreas:2012mt,Bjorken:2009mm}, NA64~\cite{Banerjee:2018vgk,Banerjee:2019hmi}, E141~\cite{Riordan:1987aw}, $\nu$-CAL~\cite{Blumlein:1990ay,Blumlein:1991xh,Tsai:2019mtm} and LHCb~\cite{LHCb:2017trq,LHCb:2019vmc}. These limits are also shown in the left panel of Fig.~\ref{kinetic_gX_limits} along with the region of the parameter space allowed by the DM relic density (light purple region). This region is obtained for a range of DM masses while still taking $g_X=0.01$. The value of $g_X$ is constrained by the Planck experiment~\cite{Ade:2015xua,Slatyer:2015jla} via the annihilation channel $D\bar{D}\to\gamma'\gamma'\to 4e$. The recast limits are shown in the right panel of Fig.~\ref{kinetic_gX_limits} for three values of $m_D$.

\section{Conclusions\label{sec7}}

While the existence of DM is now well established, its precise identification such as its mass, spin and other properties still remain unknown. In particular, the search for DM in the GeV and sub-GeV mass regimes has gained interest in view of many possible candidates in this category which include axions, sterile neutrinos and dark photons. The XENON1T results have recently indicated an excess rate in the $2\,$--$\,3\,$keV energy region of the electron recoil band, which can, among other possibilities, be explained by inelastic DM scattering processes. Here, a dark particle down scatters from a bound electron in a xenon atom to another lighter dark particle close in mass and an ejected recoil electron with excess energy equal to the mass difference of the initial heavier and final lighter dark particle. In this paper, we focused on annual modulation of a possible DM signal, which is now a standard tool utilized in experimental DM searches. We pointed out that not only the rate, but also the recoil energy exhibits an annual modulation, which is proportional to the DM mass, so that a measurement of the recoil energy modulation would allow for a direct determination of the DM mass. We proposed to analyze the annual modulation using both time and recoil energy as variables rather than time only and carried out simulations for the running conditions of the XENON1T and XENONnT experiments. We found that the new proposed technique allows for a factor 1.5 times larger sensitivity to DM detection over the conventional one. In particular, the DM mass could be determined over the projected lifespan of XENONnT with a precision of about 1\,GeV. The analysis was carried out using a dark photon model proposed recently for the analysis of the observed XENON1T signal. However, our proposed technique to use both recoil energy and time of each event in an unbinned two-dimensional fit is valid for a large class of models that might provide an explanation of the XENON1T excess or for analyses of inelastic and elastic DM detection processes in general. \\
 
\noindent
\textbf{Acknowledgments:} The research of AA, LA, MK and CW was supported by the DFG through the Research Training Network 2149 ``Strong and Weak Interactions$-$from Hadrons to Dark Matter" and the research of PN was supported in part by the United States NSF Grant PHY-1913328.

\appendix

\section{Details of the dark photon model\label{app:A}}

The model used in the analysis was proposed in ref.~\cite{Aboubrahim:2020iwb}. Here we give a 
brief summary and note the relevant formulae used in the analysis. 
The proposed model of inelastic dark matter scattering 
extends the Standard Model (SM) gauge group by an extra $U(1)_X$ under which the SM is neutral. The extra gauge field $C^{\mu}$ mixes with the SM $U(1)_Y$ hypercharge $B^{\mu}$ 
via kinetic mixing~\cite{Holdom:1985ag}.
Further, we use the Stueckelberg mechanism~\cite{Kors:2004dx}
 to generate mass for the gauge boson of the hidden sector. The total Lagrangian is then given by
\begin{align}
\mathcal{L}&=\mathcal{L}_{\rm SM}+\Delta\mathcal{L},\nonumber\\
\Delta\mathcal{L}\supset &-\frac{1}{4}C_{\mu\nu}C^{\mu\nu}-\frac{\delta}{2}C_{\mu\nu}B^{\mu\nu}+g_X J^{\mu}_X C_{\mu}-\frac{1}{2}(\partial_{\mu}\sigma+M_1 C_{\mu}+M_2 B_{\mu})^2\nonumber\\
&+\mathcal{L}_D\,,
\label{deltaL}
\end{align}
where $g_X$ is the gauge coupling in the hidden sector, $J_X$ is the hidden sector current and $\sigma$ is a pseudoscalar field which is absorbed in a gauge-invariant way via the Stueckelberg mechanism to give mass to the extra neutral gauge boson which we call $\gamma'$ (the dark photon).   
Further one may introduce matter in the hidden sector which is neutral under $U(1)_Y$ but charged under $U(1)_X$~\cite{Kors:2004dx,Cheung:2007ut}.
 More generally one may have both kinetic mixing and mass mixing~\cite{Feldman:2007wj}.
The $\mathcal{L}_D$ contains the kinetic, mass and interaction of Dirac fermions.
 It arises as follows:
we assume that the hidden sector where $C^\mu$ resides contains two mass degenerate 
 Dirac fermions $D_1$ and $D_2$ with the common mass $\mu$ 
 which, however, carry different charges $Q_1$ and $Q_2$ under 
 the $U(1)_X$ gauge group.  The interaction Lagrangian
for the hidden sector is then given by~\cite{Aboubrahim:2020iwb}
\begin{align}
 \mathcal{L}^{\rm int}_D=& -g^{\gamma'}_X Q_1 \bar D_1\gamma^\mu D_1 A_\mu^{\gamma'} - g^{\gamma'}_X Q_2 \bar D_2\gamma^\mu D_2  A_\mu^{\gamma'}-g^{Z}_X Q_1 \bar D_1\gamma^\mu D_1 Z_\mu-g^{Z}_X Q_2 \bar D_2\gamma^\mu D_2  Z_\mu,
\label{d1d2-1}
\end{align}
with $g^{\gamma'}_X =g_X(\mathcal{R}_{11}-s_{\delta}\mathcal{R}_{21})$ and $g^{Z}_X =g_X(\mathcal{R}_{12}-s_{\delta}\mathcal{R}_{22})$, where $\mathcal{R}_{ij}$ are elements of the matrix in~\cite{Aboubrahim:2020iwb,Feldman:2007wj} and $s_\delta\equiv \sinh\delta$.
 The coupling of the dark photon with the visible sector is given by
\begin{equation}
\mathcal{L}_{\rm SM}=\frac{g_2}{2\cos\theta}\bar\psi_f\gamma^{\mu}\Big[(v'_f-\gamma_5 a'_f)A^{\gamma'}_{\mu}\Big]\psi_f\,,
\label{SMLag}
\end{equation}
where $f$ runs over all SM fermions and 
\begin{equation}
\begin{aligned}
v'_f&=-\cos\psi[(\tan\psi+\bar\epsilon\sin\theta)T_{3f}-2\sin^2\theta(\bar\epsilon\csc\theta+\tan\psi)Q_f],\\
a'_f&=-\cos\psi(\tan\psi+\bar\epsilon\sin\theta)T_{3f},
\end{aligned}
\end{equation}
where $T_{3f}$ is the third component of the isospin, $Q_f$ is the electric charge and $\bar \epsilon= \epsilon \cosh \delta-\sinh \delta$. 
Three angles $\theta, \phi, \psi$ appear in the diagonalization of the $3\times 3$ gauge boson mass matrix~\cite{Feldman:2007wj} and are defined so that
\begin{align}
\tan\theta &= \frac{g_Y}{g_2} \cosh\delta \cos\phi, \quad \tan\phi= \bar \epsilon, \quad 
\tan 2\psi\simeq \frac{2 \, \bar \epsilon \,m_Z^2 \sin\theta}{ m_{\gamma'}^2 
-m_Z^2 +(m_{\gamma'}^2+m_Z^2 -m_W^2) \bar\epsilon^2}. 
\label{angles}
 \end{align}
 The angle $\theta$ that appears in Eq.~(\ref{SMLag}) is essentially the weak angle
 in the extended model and in the standard model limit it is defined by $\tan\theta=g_Y/g_2$
 since $\delta, \phi \to 0$ in that limit as can be seen from Eq.~(\ref{angles}).
 
 To generate inelastic scattering we need to split the masses of the $D$-fermions. To this end we
add a $U(1)_X$ gauge violating mass terms $\Delta\mu (\bar D_1D_2 + \text{h.c.})$ so that the Lagrangian for the
 $(D_1, D_2)$ mass terms  is  given by 
 \begin{align}
  \mathcal{L}^{\rm mass}_D = 
 - \mu (\bar D_1 D_1 + \bar D_2 D_2)-
 \Delta \mu (\bar D_1 D_2 + \bar D_2 D_1).
 \label{d1d2-3}
 \end{align}
We can now go to the mass diagonal basis with Dirac fermions $D_1'$ with mass $m_1=\mu-\Delta \mu$
and $D_2'$ with mass $m_2= \mu+\Delta \mu$ and we assume $m_2>m_1$ so $D_2'$ is the heavier
of the two dark fermions. 

\section{Details of the inelastic $D_2'(\vec{p}_2)+e(\vec{p}_1)\to D_1'(\vec{p}_4)+e'(\vec{p}_3)$ cross-section\label{app:B} }
  
Here we give details of the inelastic scattering cross-section of the process described $D_2'(\vec{p}_2)+e(\vec{p}_1)\to D_1'(\vec{p}_4)+e'(\vec{p}_3)$. Assuming the dark photon mass is much greater than the momentum transfer, the averaged matrix element squared for this process
 is given by
\begin{align}
\overline{|\mathcal{M}|^2}&=\frac{2\bar{g}_X^2 g_2^2}{m^4_{\gamma'}\cos^2\theta}\Bigg\{\frac{1}{2}(a_f^{\prime2}-v_f^{\prime2})\Big[(m_1-m_2)^2-(t+2m_1m_2)\Big]m^2_e +\frac{1}{4}(v_f^{\prime2}+a_f^{\prime2})\Big[(m_2^2+m_e^2-u) \nonumber \\
&\times (m_1^2+m_e^2-u)+(s-m_1^2-m_e^2)(s-m_2^2-m_e^2)-2m_1m_2(2m_e^2-t)\Big]\Bigg\},
\end{align} 
where $\bar{g}_X=\frac{1}{2} g_X(Q_1-Q_2)$ and $s,t,u$ are the Mandelstam variables.
The directional matrix element for free electron-DM scattering is given by
\begin{equation}
\overline{|\mathcal{M}(\vec{q})|^2}=\overline{|\mathcal{M}(q)|^2}\times |F_{DM}(q)|^2,
\end{equation}
where the form factor $F_{DM}(q)$ can be taken as 1 for a small momentum transfer. The electron-DM scattering differential cross-section is given by
\begin{align}
\frac{d\bar{\sigma}_e}{d\Omega}&=\frac{1}{64\pi^2 s}\frac{|\vec{p}_3|}{|\vec{p}_1|}\overline{|\mathcal{M}(\vec{q})|^2}.
\end{align}
Keeping the velocity dependence in the Mandelstam variables, we have
\begin{equation}
t=-q^2\simeq -\Delta m\left(1-\sqrt{\frac{2m_e v^2}{\Delta m}}\cos\theta_{\rm CM}\right),~~~s\sim (m_2+m_e)^2\left(1+\frac{\mu_{De}}{m_2+m_e}v^2\right),
\end{equation}
and integrating over $\theta_{\rm CM}$, the scattering angle in the CM frame, and over $\phi$, we get for the DM-e scattering cross-section
\begin{equation}
\bar{\sigma}_e\simeq\frac{\bar{g}_X^2 g_2^2}{16\pi m^4_{\gamma'}\cos^2\theta}\left(\frac{4\mu^2_{De}}{1+\frac{\mu_{De}}{m_2+m_e}v^2}\right)\left[v^{\prime 2}_f+(a_f^{\prime 2}+v^{\prime 2}_f) v^2\right],
\end{equation} 
where $\mu_{De}=\frac{m_2 m_e}{m_2+m_e}$ is the dark matter-electron reduced mass. For $v\sim 10^{-3}$, one can discard the velocity-dependent piece which gives the cross-section of Eq.~(\ref{inelastic-cross}).

\section{Approximate equations for the event rate\label{app:C}}

Knowing that $|\mathbf{v}+\mathbf{v}_{EH}|<v_{\rm esc}$ and that $v_{\rm esc}>v_{EH}$, we have
\begin{align}
\eta(E,t)&=\int_{|\mathbf{v}+\mathbf{v}_{EH}|>v_{\rm min}}\frac{f(\textbf{v}+\textbf{v}_{EH})}{v}d^3 v \nonumber \\
&=\frac{1}{2 N_{\rm esc} v_{EH}(t)}\left[\text{Erf}\left(\frac{v_{EH}(t)+v_{\rm min}(t)}{v_0}\right)+\text{Erf}\left(\frac{v_{EH}(t)-v_{\rm min}(t)}{v_0}\right)-\frac{4v_{EH}(t)}{\sqrt{\pi}v_0}e^{-v^2_{\rm esc}/v^2_0}\right].
\end{align} 
In the limit $v_{ES}\ll v_{SH},v_{EH}$, we can rewrite Eq.~(\ref{relv}) as $v_{EH}(t)\simeq u+\Theta(t)$, where
\begin{equation}
\begin{aligned}
u&=(v^2_{ES}+v^2_{SH})^{1/2}, \\
\Theta(t)&=\frac{v_{ES}v_{SH}}{2(v^2_{ES}+v^2_{SH})^{1/2}}\cos(2\pi t-\phi).
\end{aligned}
\label{appr}
\end{equation}
For $v_{\rm esc}\rightarrow\infty$, $N_{\rm esc}\rightarrow 1$ and using Eq.~(\ref{appr}), we expand Eq.~(\ref{etaf}) in $\Theta(t)$ to get
\begin{equation}
\eta(E',t)\simeq P(E')+Q(E')\cos(2\pi t-\phi),
\end{equation}
where
\begin{align}
P(E')&=\frac{1}{2u}\left[\text{Erf}\left(\frac{u-v_{\rm min}(E')}{v_0}\right)+\text{Erf}\left(\frac{u+v_{\rm min}(E')}{v_0}\right)\right], \\
Q(E')&=\frac{v_{ES}v_{SH}}{2 u^2}\left[\frac{1}{\sqrt{\pi}v_0}\left(e^{-(u-v_{\rm min})^2/v^2_0}+e^{-(u+v_{\rm min})^2/v^2_0}\right)-P(E')\right].
\end{align}
The event rate is given by
\begin{align}
\frac{dR}{dE\,dt}(E,t)&=n_{\textrm{Xe}} \frac{\rho_2}{m_2} \frac{\bar\sigma_e}{2m_e} \int  \eta(E_v,t) \, K'(E_v, t) \, R_S(E, E_v)\,dE_v \nonumber \\
&\simeq W(E,t)+Z(E,t)\cos(2\pi t-\phi),
\end{align}
where
\begin{align}
W(E,t)&=n_{\textrm{Xe}} \frac{\rho_2}{m_2} \frac{\bar\sigma_e}{2m_e} \int P(E_v) \, K'(E_v, t) \, R_S(E, E_v)\,dE_v\,,  \\
Z(E,t)&=n_{\textrm{Xe}} \frac{\rho_2}{m_2} \frac{\bar\sigma_e}{2m_e} \int Q(E_v) \, K'(E_v, t) \, R_S(E, E_v)\,dE_v\,.
\end{align}

\end{document}